\documentclass[sigplan,10pt,nonacm]{acmart}

\usepackage{bibunits}
\usepackage{graphicx}
\usepackage{xcolor}
\usepackage{listings}
\usepackage{xspace}
\usepackage{amsmath,amsfonts, setspace}
\usepackage{algpseudocode}
\usepackage{enumitem}
\usepackage{multirow}
\usepackage{tabularx, booktabs}
\usepackage{subcaption}
\usepackage{makecell}
\usepackage[scaled=0.85]{beramono}
\usepackage{wrapfig}
\usepackage{bm}
\usepackage{balance}
\usepackage{pifont} 

\usepackage[ruled,linesnumbered]{algorithm2e}
\usepackage{amsthm}
\definecolor{dkgreen}{rgb}{0,0.6,0}
\definecolor{gray}{rgb}{0.5,0.5,0.5}
\definecolor{mauve}{rgb}{0.58,0,0.82}
\definecolor{awesome}{rgb}{1.0, 0.13, 0.32}
\definecolor{beaver}{rgb}{0.62, 0.51, 0.44}
\definecolor{carrotorange}{rgb}{0.93, 0.57, 0.13}
\definecolor{chocolate}{rgb}{0.82, 0.41, 0.12}
\definecolor{copper}{rgb}{0.72, 0.45, 0.2}
\definecolor{crimsonglory}{rgb}{0.75, 0.0, 0.2}

\newcommand{\clz}[1]{{\color{blue}{\small (clz: [#1])}}}

\newcommand{\todo}[1]{{\color{red}{\xspace #1}}}

\newif\ifrevision
\revisionfalse

\ifrevision
    \newcommand{\rev}[1]{\textcolor{red}{#1}}
\else
    \newcommand{\rev}[1]{#1}
\fi

\theoremstyle{definition}

\lstdefinestyle{base}{
  language=C,
  emptylines=1,
  breaklines=true,
  basicstyle=\ttfamily\color{blue},
  moredelim=**[is][\color{red}]{@}{@},
}
\newcommand{\projectname}{{GANNS}\xspace}
\usepackage[normalem]{ulem}

\usepackage{etoolbox}

\SetCommentSty{mycommentstyle}

\copyrightyear{2027}
\acmYear{2027}
\setcopyright{cc}
\setcctype{by}

\acmConference[EuroSys '27]
{22nd European Conference on Computer Systems}
{April 19--23, 2027}
{Rabat, Morocco}

\acmBooktitle{22nd European Conference on Computer Systems
(EuroSys '27), April 19--23, 2027, Rabat, Morocco}

\acmDOI{10.1145/3842654.3848536}
\acmISBN{979-8-4007-2971-3/2027/04}

\renewcommand\footnotetextcopyrightpermission[1]{}
\begin{document}


\title{Beyond Overlap: Hierarchical Graph Abstraction for Out-of-Core Vector Search}


\author{\texorpdfstring{
Lizheng Chen$^{1,\dagger}$,
Pinhuan Wang$^{1,\dagger}$,
Shaonan Ma$^{2,\dagger}$,
Jie Zhang$^{3}$,
Qing Wang$^{1}$,
Zhengyi Yang$^{4}$,\\
Heng Zhang$^{5}$,
Mingxing Zhang$^{6}$,
Fan Yang$^{7}$,
Guihai Chen$^{1}$,
Chen Tian$^{1}$,
Chengying Huan$^{1,*}$\\[4pt]
$^{1}$Nanjing University \quad
$^{2}$Qiyuan Lab \quad
$^{3}$Peking University \quad
$^{4}$University of Sydney \\
$^{5}$Institute of Software, Chinese Academy of Sciences \quad
$^{6}$Tsinghua University \quad
$^{7}$Huawei
}{Lizheng Chen, Pinhuan Wang, Shaonan Ma, Jie Zhang, Qing Wang, Zhengyi Yang, Heng Zhang, Mingxing Zhang, Fan Yang, Guihai Chen, Chen Tian, Chengying Huan}}
\renewcommand{\shortauthors}{L. Chen, P. Wang et al.}

\gdef\authors{%
Lizheng Chen\and
Pinhuan Wang\and
Shaonan Ma\and
Jie Zhang\and
Qing Wang\and
Zhengyi Yang\and
Heng Zhang\and
Mingxing Zhang\and
Fan Yang\and
Guihai Chen\and
Chen Tian\and
Chengying Huan%
}

\begin{abstract}
At billion scale, approximate nearest neighbor search (ANNS) often becomes an out-of-core problem: the full vector collection and index structures exceed memory capacity, making query performance dominated by SSD accesses and DRAM-SSD data movement.
Existing systems struggle to strike a balance between accuracy and efficiency: physical-overlap methods replicate vectors or index entries across partitions, enlarging the SSD-resident index and incurring extra I/O; quantization-based methods reduce memory usage, but their approximate distances are less reliable and often require costly raw-vector reranking from SSD to preserve recall.

We present \projectname{} (\textbf{G}raph-Abstraction ANNS), a hierarchical out-of-core ANNS engine that improves accuracy and I/O efficiency simultaneously. 
\projectname{} stores each cluster as a disjoint SSD partition with scale-aware indexes, while a memory-resident graph abstraction provides logical overlap before SSD access.
During serving, \projectname{} uses query hotness and cluster priorities from the graph abstraction to prune low-value clusters and improve access locality. 
\rev{Across five datasets under strict memory constraints, \projectname delivers representative QPS improvements of 2.0--4.7$\times$ over state-of-the-art baselines, while preserving accuracy.}
\end{abstract}



\maketitle

\begingroup
\renewcommand{\thefootnote}{}
\footnotetext{\textsuperscript{*}Corresponding author.\quad
\textsuperscript{$\dagger$}Equal contribution.}
\endgroup

\section{Introduction}
\label{sec:intro}
Approximate nearest neighbor search (ANNS) is a core primitive in modern AI and data-intensive systems, including vector databases~\cite{wang2021milvus,pan2024vector}, recommendation engines~\cite{covington2016deep,okura2017embedding,Pal2020PinnerSage}, web search~\cite{zhang2022uni}, and large language model (LLM) inference~\cite{liu2024retrievalattention,zhang2025pqcache}. 
In particular, LLM services such as ChatGPT~\cite{roumeliotis2023chatgpt} and Claude~\cite{adetayo2024microsoft} rely on vector search for retrieval-augmented generation (RAG)~\cite{lewis2020retrieval,cuconasu2024power,Seemakhupt2024EdgeRAGOR}, context reuse~\cite{yao2025cacheblend,Agarwal2025cachecraft}, agentic or knowledge-intensive retrieval~\cite{hu2025hedrarag}, and sparse-attention serving~\cite{deng2025alayadb,chen2024magicpig}.
In all these settings, ANNS retrieves a small set of vectors relevant to a query embedding, enabling downstream systems to operate on query-relevant data without exhaustive scans over massive vector collections.

As vector collections grow to billion scale, raw vectors and high-quality indexes can exceed available DRAM capacity, making ANNS an out-of-core problem. 
Out-of-core ANNS therefore places vectors and most index structures on SSDs, while keeping only compact routing or filtering metadata in memory~\cite{jayaram2019diskann,chen2021spann,wang2024starling,guo2025achieving,kang2025scalable,chen2026alayalaser}. 
This setting is especially relevant to LLM services, where vector search has become a critical component for retrieval and context management; production-style systems such as OpenAI Vector Store~\cite{openai} expose vector storage as a first-class service for large document collections. 
At the same time, GPU HBM in these services is primarily consumed by model-side resources such as LLM weights, KV cache, and activation buffers.
Some recent systems explore CPU--GPU--SSD collaboration for billion-scale ANNS~\cite{tian2025towards,jiang2025high}.
However, they assume that GPU HBM can be provisioned for vector search, which may be difficult in RAG and agentic-AI serving~\cite{lewis2020retrieval,yao2022react,jin2024ragcacheefficientknowledgecaching,wu2023autogenenablingnextgenllm}. 
This paper therefore targets the widely deployable CPU--SSD out-of-core setting, where DRAM is reserved for compact guidance and runtime metadata, while large vectors and local indexes remain SSD-resident.

Existing CPU–SSD out-of-core ANNS systems are mainly organized as graph-based or partition-based indexes. 
Graph-based systems such as DiskANN~\cite{jayaram2019diskann}, Starling~\cite{wang2024starling}, and PipeANN~\cite{guo2025achieving} store most graph records on SSD and reduce random I/O through memory-resident metadata, graph reordering, or pipelined traversal. 
Partition-based systems such as SPANN~\cite{chen2021spann} divide vectors into Inverted File (IVF) posting lists or clusters~\cite{jegou2011productSIFT}, search only a subset of promising partitions, and often rely on an in-memory routing layer to select them. 
Both approaches make billion-scale search feasible under limited DRAM, but high-recall searches still require boundary reachability: the ability to reach neighbors separated by graph, shard, or partition boundaries.

Existing designs commonly support boundary reachability through either physical overlap in the SSD index or quantized search with raw-vector reranking.
For graph-based indexes, physical overlap is realized through graph connectivity: DiskANN builds local graphs from overlapping build partitions and merges their edges into a global SSD-resident graph; at query time, reachability is provided by SSD-resident edges rather than vector replicas.
For partition-based indexes, physical overlap is realized through replication: SPANN places boundary vectors in multiple posting lists so that a query can reach them with only a few probes.
These mechanisms improve reachability but often increase out-of-core overhead through random graph page accesses, replicated entries, or larger posting lists.
Another approach is quantized search with raw-vector reranking: systems use Product Quantization (PQ)~\cite{jegou2011productSIFT} or Inverted File with Product Quantization (IVFPQ)~\cite{jegou2011productSIFT} to maintain compact representations in memory, score candidates approximately, and then fetch full vectors from SSD for exact reranking~\cite{jayaram2019diskann,tian2025towards}.
This reduces resident memory, but high-recall search still depends on costly raw-vector reranking, turning quantization error into additional SSD traffic.

\begin{figure}[!t]
    \vspace{-0.8em}
    \centering
    \includegraphics[width=.85\linewidth]{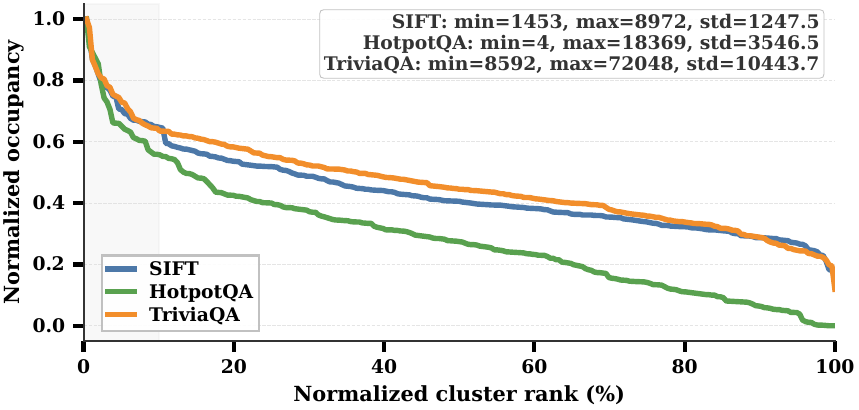}
    \vspace{-0.6em}
    \caption{Skewness on SIFT, HotpotQA, and TriviaQA.}
    \label{fig:skewness}
    \vspace{-1em}
\end{figure}

\noindent\textbf{Design Principle: locality-preserving reachability.}
The limitation above exposes a framework question: high recall needs boundary reachability, yet placing this reachability in the SSD layout or relying on raw-vector reranking introduces extra SSD work.
The system must preserve reachability to neighbors across partition or storage boundaries without sacrificing local and bounded SSD accesses.

Our measurements suggest that disjoint SSD partitions with partition-local indexes provide a better out-of-core substrate.
This substrate preserves physical locality: each vector remains in one SSD partition, and SSD search touches only selected partition-local data.
As described in Section~\ref{sec:motivation-efficiency}, one instantiation with graph-style local indexes preserves coarse I/O locality while avoiding exhaustive scans.


\noindent\textbf{Our design.}
\rev{The architectural insight behind our design is to \textit{decouple physical locality from logical connectivity}.
A compact memory-resident graph abstraction represents each partition with a stable centroid and selected representative vectors, exposing partition-level evidence before SSD access.
Rather than serving only as a routing layer, this partition-level evidence is reused to rank and prune candidate partitions and to organize query execution for better locality.
The SSD layer then preserves physical locality by searching only selected non-replicated partitions with scale-aware local search structures.}

\noindent\textbf{Challenge I: scale-aware partition search.}
Once boundary recovery is moved to memory, the remaining question is how to search selected SSD partitions efficiently under skewed cluster sizes.
As shown in Figure~\ref{fig:skewness}, IVF partitioning produces long-tailed clusters even on SIFT (1,453--8,972 vectors, std. 1,247.5) and becomes much more skewed on TriviaQA (8,592--72,048 vectors, std. 10,443.7).
Such skewness makes a fixed local index inefficient: tiny partitions may be cheaper to scan than to index, while large partitions can impose high metadata and memory pressure, as quantified in Section~\ref{sec:independent-partitions}.
This calls for stable SSD partitions with local access paths adapted to partition scale and memory pressure.

\noindent\textbf{Challenge II: utility-aware out-of-core execution.}
Beyond choosing how to search each selected partition, an out-of-core engine must also decide which partition accesses are worth issuing.
The memory-resident graph abstraction exposes partition-level signals before SSD access, and serving workloads repeatedly visit a small subset of partitions, as shown in Figure~\ref{fig:hot-cluster}.
However, frequently visited partitions are not necessarily useful for every query, and blindly searching them can still waste SSD I/O.
Coarse partition-level signals must therefore be translated into fewer unnecessary out-of-core accesses without sacrificing recall.
The challenge is to distinguish useful partition accesses from redundant ones before local search and to organize the remaining accesses for better page reuse.

We present \projectname{} (\textbf{G}raph-Abstraction ANNS), a hierarchical out-of-core ANNS engine that decouples physical locality from logical connectivity.
To preserve locality while maintaining reachability, \projectname{} stores vectors in disjoint SSD partitions and maintains a graph abstraction composed of centroid anchors and representative vectors.
To handle skewed partitions, the SSD layer searches partitions with local indexes adapted to partition scale and memory pressure.
To reduce redundant out-of-core work, \projectname{} uses per-query partition signals to refresh the graph abstraction, prioritize useful partitions, prune low-value accesses, and schedule queries with similar partition visits close together.
Together, these mechanisms improve routing precision, reduce redundant SSD searches, and increase temporal locality.

This paper makes the following contributions:

\begin{itemize}[leftmargin=*,topsep=0pt,itemsep=0pt,parsep=0pt]
  \item \textbf{A decoupled out-of-core ANNS architecture.}
  We identify that existing high-recall mechanisms incur extra SSD cost: physical overlap enlarges the SSD-resident index or increases random graph accesses, while quantized search requires full-vector SSD reads for reranking.
  \rev{We decouple physical locality from logical connectivity: vectors remain in non-replicated SSD partitions, while a memory-resident graph abstraction provides boundary reachability and exposes reusable partition-level evidence before SSD access.}

  \item \textbf{Shape- and scale-aware partition search.}
  We design \projectname{}'s hierarchical search structure, where a compact graph abstraction with representative vectors exposes partition-level evidence before SSD access. 
  The SSD layer then searches selected partitions using local indexes adapted to partition scale and memory pressure.

  \item \textbf{Access-hotness-aware execution.}
  We design runtime optimizations that turn partition-level access hotness into out-of-core gains. 
  \projectname{} refreshes the graph abstraction toward frequently useful regions, prioritizes and prunes candidate partitions using per-query graph evidence, and schedules queries with similar partition visits to improve reuse among remaining SSD accesses.

  \item \textbf{Implementation and evaluation.}
  We implement \projectname as a CPU--SSD ANNS engine and evaluate it against representative graph-based, partition-based baselines.
  \rev{Under memory constraints, \projectname achieves representative QPS improvements of 2.0--4.7$\times$ while preserving recall.}
\end{itemize}

\section{Background}
\label{sec:background}

\subsection{Out-of-Core ANNS Architectures}
\label{sec:bg-architectures}

\begin{figure}[t]
    \centering
    \vspace{-0.4em}
    \begin{subfigure}[t]{0.4\linewidth}
        \centering
        \includegraphics[width=\linewidth]{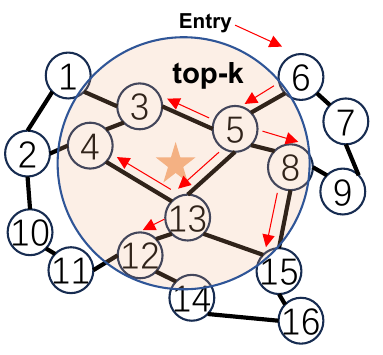}
        \caption{Graph-based index.}
        \label{fig:bg-graph}
    \end{subfigure}
    ~~~~~~~~~~
    \begin{subfigure}[t]{0.4\linewidth}
        \centering
        \includegraphics[width=\linewidth]{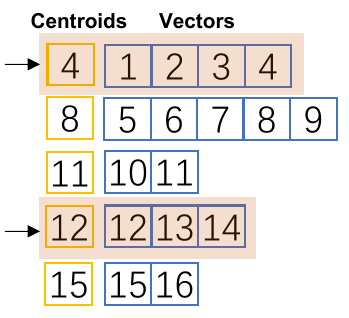}
        \caption{Partition-based index.}
        \label{fig:bg-partition}
    \end{subfigure}
    \vspace{-0.4em}
    \caption{Two basic out-of-core ANNS organizations. A graph index searches by traversing a proximity graph from an entry point toward the query region, while a partition index first selects promising partitions and then searches only the corresponding posting lists or clusters.}
    \label{fig:bg-architectures}
    \vspace{-0.4em}
\end{figure}

ANNS systems reduce the search space mainly through \emph{graph-based} or \emph{partition-based} indexing. 
In a graph-based index~\cite{jayaram2019diskann,wang2024starling,guo2025achieving,chen2026alayalaser,kang2025scalable}, vectors are organized as a proximity graph, and search starts from one or more entry points and greedily traverses graph edges toward the query-relevant region, as illustrated in Figure~\ref{fig:bg-graph}. 
DiskANN~\cite{jayaram2019diskann} is a representative out-of-core graph system: it keeps PQ-compressed vector representations in DRAM for distance estimation, while storing the graph and full-precision vectors on SSD. 
This design preserves high recall, but SSD-resident graph traversal may still introduce many random page accesses.

A partition-based index~\cite{jegou2011productSIFT,chen2021spann,zhang2024fast,tian2025towards} instead divides the dataset into clusters or posting lists, typically using IVF or $k$-means, and searches only a subset of partitions per query. 
As shown in Figure~\ref{fig:bg-partition}, the system scans compact partition summaries, ranks partitions by their distance to the query, and searches the nearest vector lists. 
Systems such as SPANN~\cite{chen2021spann} and RUMMY~\cite{zhang2024fast} follow this direction. 
This design naturally bounds the out-of-core search footprint, but partitioning also weakens connectivity across partition boundaries: true neighbors of a query may be split across multiple partitions.


Some graph-based out-of-core systems also use a small memory-resident graph routing layer above SSD-resident data~\cite{wang2024starling, guo2025achieving}. 
\rev{Such in-memory routing layers primarily provide query-near entry
points for subsequent SSD-resident graph traversal, rather than explicitly
representing logical overlap among disjoint physical partitions.}
Existing routing nodes are often centroids~\cite{tian2025towards} or sampled vectors~\cite{wang2024starling}. 
However, centroids are weak at describing off-center regions in high-dimensional partitions, while sampled vectors may provide closer graph entries but are noisy for partition-level decisions and can mislead the system toward less useful SSD regions (Section~\ref{sec:component-ablation}).

\subsection{Physical Overlap and Quantized Search}
\label{sec:bg-overlap-compression}

A central challenge in out-of-core ANNS is how to preserve accuracy when true neighbors are separated by graph, shard, or partition boundaries. 
Existing systems usually address this problem through the following two mechanisms.

\begin{figure}[t]
    \centering
    \vspace{-0.4em}
    \begin{subfigure}[t]{0.31\linewidth}
        \centering
        \includegraphics[width=\linewidth]{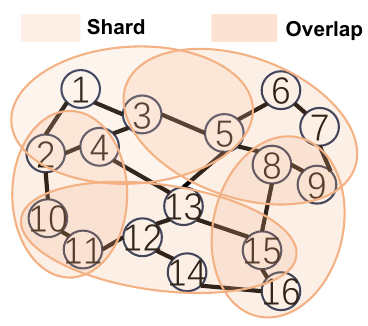}
        \caption{Graph-based physical overlap.}
        \label{fig:bg-overlap-graph}
    \end{subfigure}
    \hfill
    \begin{subfigure}[t]{0.31\linewidth}
        \centering
        \includegraphics[width=\linewidth]{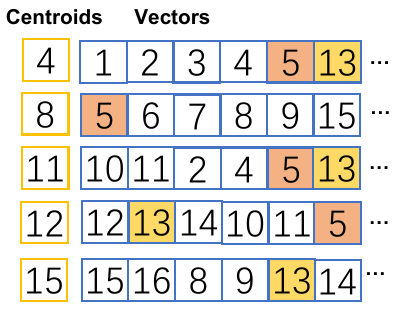}
        \caption{Partition-based physical overlap.}
        \label{fig:bg-overlap-partition}
    \end{subfigure}
    \hfill
    \begin{subfigure}[t]{0.31\linewidth}
        \centering
        \includegraphics[width=\linewidth]{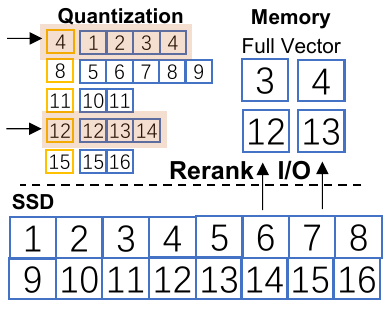}
        \caption{Quantized search with reranking.}
        \label{fig:bg-rerank}
    \end{subfigure}
    \vspace{-0.4em}
    \caption{Existing mechanisms for supporting high recall in out-of-core ANNS.
    Figures~\ref{fig:bg-recovery}a and~\ref{fig:bg-recovery}b encode boundary reachability in the SSD-resident index through graph connectivity or boundary-vector replication, while Figure~\ref{fig:bg-recovery}c stores quantized representations in memory and fetches full vectors from SSD for reranking.}
    \label{fig:bg-recovery}
    \vspace{-0.4em}
\end{figure}

The first direction is physical overlap.
In graph-based systems, physical overlap may appear during construction. 
For example, DiskANN~\cite{jayaram2019diskann} builds a large SSD-resident graph from overlapping partitions: a vector can participate in multiple build partitions, local graphs are constructed under memory limits, and their edges are then merged into a global graph. 
At search time, cross-boundary reachability is preserved by SSD-resident graph edges, as illustrated in Figure~\ref{fig:bg-overlap-graph}. 
This helps recover neighbors across storage regions, but may also lead to random accesses across distant SSD pages. \rev{Thus, DiskANN encodes cross-region reachability in SSD-resident graph edges, which queries must traverse at search time.}
In partition-based systems, overlap is more explicit in the index layout.
As shown in Figure~\ref{fig:bg-overlap-partition}, SPANN~\cite{chen2021spann} replicates boundary vectors into multiple posting lists, so that a query can find neighbors across partition boundaries even when it probes only a few lists.
Such replication improves boundary robustness, but it also enlarges posting lists and increases the amount of SSD-resident data that must be stored. \rev{Thus, SPANN encodes cross-partition reachability by duplicating boundary vectors across SSD posting lists.}

The second direction is quantized search with reranking.
As shown in Figure~\ref{fig:bg-rerank}, compact PQ codes are kept in memory for approximate scoring, while selected full vectors are fetched from SSD for exact reranking.
This approach reduces resident memory, but quantized distances are lossy. 
Therefore, high-recall search often still requires raw-vector reranking, which introduces additional SSD reads.

\subsection{Limitations of Prior Works}
\label{sec:bg-limitations}


The mechanisms in Figure~\ref{fig:bg-recovery} support high recall, but they add work to the out-of-core layer.
Physical overlap places boundary reachability in the SSD-resident index, which leads to replicated entries, larger posting lists, or random graph accesses.
Quantized search keeps the resident index compact, but approximate quantized scores still require many raw-vector reads from SSD for exact reranking at high recall.

We quantify this tradeoff on DEEP100M under Recall@10 metric.
The two mechanisms show different failure modes.
For SPANN, higher recall incurs costs along two dimensions. With a fixed replication factor of 4, increasing the search budget raises Recall@10 from 0.928 to 0.982, while page accesses increase from 584 to 1{,}905 per query, or $3.2\times$.
\rev{With search parameters fixed, increasing the replication factor from 4 to 8 grows the posting file from 148.78~GB to 262.97~GB (76.8\%) and increases page accesses by 47.7\%, while Recall@10 improves from 0.9280 to 0.9665, or 3.85 percentage points.}
Thus, physical overlap increases both index footprint and the SSD work needed to exploit additional boundary replicas.
For IVFPQ, quantized search has the opposite limitation: increasing \texttt{nprobe} from 8 to 128 brings a small Recall@10 improvement, from 0.411 to 0.436, while reducing QPS by $10.4\times$.
To reach high recall, IVFPQ must fetch raw vectors for reranking; reranking 1{,}000 vectors/query reaches 0.982--0.991 Recall@10, but introduces hundreds to more than one thousand page-level reads for reranking per query.


These observations point to a different design direction:
preserving high recall without enlarging SSD partitions or issuing large amounts of full-vector reads for reranking.

\section{Motivation and Design Rationale}
\label{sec:motivation}
The discussion in Section~\ref{sec:bg-limitations} shows that existing designs maintain high recall in two costly ways: physical overlap encodes boundary reachability in the SSD-resident index, while quantized search relies on raw-vector reranking.
This raises two questions: how can boundary reachability be represented efficiently, and how should partition-local search be organized under skewed workloads?
Once the SSD layer is partitioned, we refer to boundary reachability as cross-partition connectivity: reaching relevant partitions even when true neighbors are split across partition boundaries.


\subsection{Placement of Cross-Partition Connectivity}
\label{sec:motivation-efficiency}

We begin with the placement of cross-partition connectivity. In an
out-of-core setting, encoding such connectivity directly in the
SSD-resident layout can weaken partition locality, while relying only on partition-local search makes boundary-crossing neighbors difficult to recover.

To understand this tradeoff, we examine three architectures under out-of-core memory limits: partition-only IVF search, a monolithic graph-based layout~\cite{malkov2020hnsw}, and a disjoint-partition layout with local graph indices inside each partition. 
The detailed I/O breakdown is reported later in Section~\ref{sec:component-ablation}. 
On SIFT, the disjoint-partition local-index layout reduces page misses to 38.4/query under a 90\% memory limit, compared with 1009.4 for IVF and 498.4 for the monolithic graph. 
Under a tighter 50\% limit, it still lowers misses to 132.0/query, compared with 1865.6 and 1392.1, respectively. 
These results suggest that partition locality should be combined with partition-local indexing, but not through a monolithic disk-resident graph or physical replication across partitions.

Instead, the physical layout should remain partitioned and non-replicated, while cross-partition connectivity is represented logically in memory. 
This separates two roles: the memory-resident layer identifies promising partitions, and the SSD-resident layer performs local search within them.

\noindent\textbf{Implication 1: logical overlap over non-replicated partitions.}
High recall should not require duplication or full-vector reranking.
Boundary reachability should be provided through memory-resident logical overlap over non-replicated partitions, with local search adapted inside each partition.

\subsection{Shape- and Scale-Aware Search}
\label{sec:motivation-decoupling}

Once cross-partition connectivity is represented in memory, the next question is how this separation remains effective under skewed data. The memory-resident layer must provide reliable partition-level guidance before SSD access, while the SSD-resident layer must search partitions
with very different sizes and access costs without repartitioning.

\noindent\textbf{Shape-aware guidance.}
A compact memory-resident routing layer cannot rely only on centroids, as in prior designs~\cite{tian2025towards}. 
Although centroids are small and stable, they provide only a coarse summary of each partition. 
In our measurement on the largest TriviaQA cluster, most vectors are not close to the centroid: their distances to the centroid are mainly concentrated between 0.43 and 0.48, with the peak around 0.45, and only a small fraction of vectors appear below 0.40. 
This indicates that dense semantic partitions can have large regions far from the centroid, where relevant query results may still appear. 
Random samples~\cite{wang2024starling} can cover more such regions, but they are query agnostic and may spend routing capacity on areas that rarely help future queries. \rev{We use the additional representative budget to capture diverse off-center regions that complement the centroid.}


\begin{figure*}[t!]
    \centering
    \includegraphics[width=\textwidth]{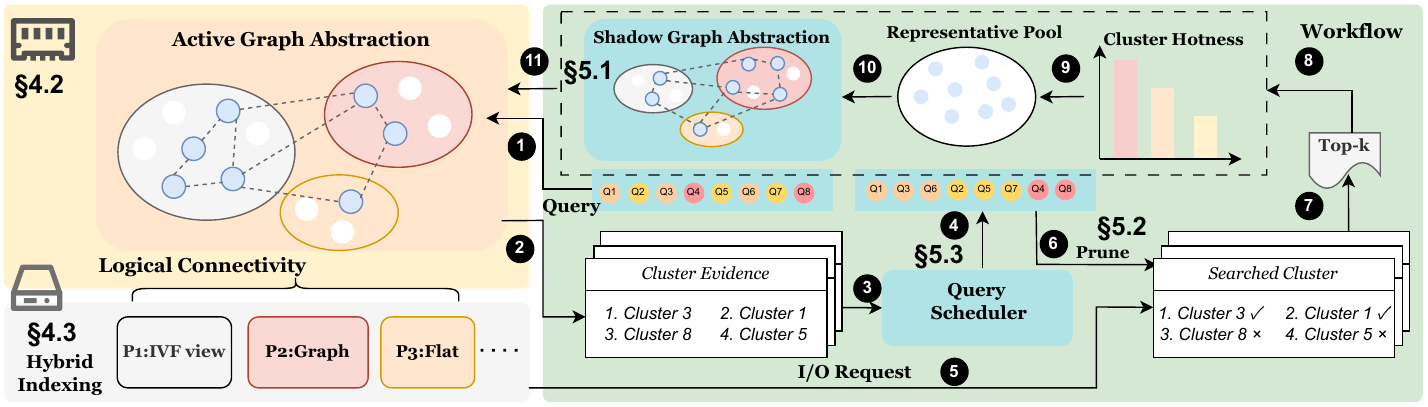} 
    \caption{Overview of \projectname, showing its hierarchical structure and query-processing workflow.}
    \label{fig:workflow}
\end{figure*}
\noindent\textbf{Implication 2: shape-aware partition guidance.}
The memory-resident routing layer should keep centroid anchors while adding representative vectors that capture off-center partition shape
before SSD access.

\noindent\textbf{Scale-aware local search.}
After the memory-resident layer identifies candidate partitions, the system must decide how to search each partition efficiently. Figure~\ref{fig:skewness} shows that IVF partitioning produces long-tailed cluster sizes across feature and semantic workloads. Even on SIFT~\cite{jegou2011productSIFT}, cluster sizes range from 1{,}453 to 8{,}972 vectors (std.\,$=1{,}247.5$). The skewness becomes more pronounced on semantic QA workloads: with \texttt{BGE} embeddings~\cite{xiao2024c}, HotpotQA~\cite{yang2018hotpotqa} contains
clusters ranging from a few vectors to more than 18K vectors, while TriviaQA~\cite{triviaqa2017} contains clusters ranging from thousands to more than 70K vectors. This long-tailed distribution means that out-of-core ANNS must serve partitions with different search costs, metadata footprints, and page-cache behavior.


\rev{A natural alternative is to adapt the physical partitioning to
evolving query skew, as explored by Quake~\cite{mohoney2025quake}.}
In an SSD-resident setting, repartitioning requires moving vectors, rewriting local index files, and rebuilding affected structures, causing large write amplification and disrupting serving.
For example, rebalancing only 1\% of a billion-scale dataset already requires rewriting millions of vectors and can take more than five hours in our storage setup.
\rev{In the host--GPU setting, RUMMY~\cite{zhang2024fast} instead addresses uneven cluster execution through cluster balancing and reordered pipelining to improve GPU utilization and overlap PCIe transfers with computation.}

\noindent\textbf{Implication 3: stable partitions with adaptive local search.}
Partition boundaries should remain stable on SSD, while the local search
structure for each partition adapts to its scale and memory pressure.

\subsection{Design Implications}
\label{sec:motivation-summary}

The observations above lead to the hierarchical design of \projectname. First, \projectname preserves physical locality by keeping SSD partitions non-replicated, and provides logical connectivity through the memory-resident graph abstraction. Second, its routing layer combines centroid anchors with representative vectors to capture partition shape. Third, its SSD-resident layer keeps partition boundaries fixed but adapts local access paths to partition scale. These requirements motivate the architecture in Section~\ref{sec:architecture}, while the cluster evidence exposed by this architecture further enables the execution-layer optimizations in Section~\ref{sec:system-optimization}.

\section{\projectname: Hierarchical Out-of-Core ANNS}~\label{sec:architecture}

\vspace{-1.0em}
\subsection{\projectname Overview}
\label{sec:arch-workflow}

To improve the accuracy-memory tradeoff, we introduce \projectname, a hierarchical out-of-core ANNS engine that decouples physical locality from logical connectivity. Figure~\ref{fig:workflow} illustrates the overall workflow. \projectname first partitions the vector set into disjoint clusters with $k$-means, assigning each vector to exactly one physical partition. It then builds a compact memory-resident graph abstraction $\mathcal{GA}$ using cluster centroids and selected data proxies, which characterize each cluster from multiple directions and
radii (Section~\ref{sec:dynamic-navigation}). Finally, \projectname constructs an independent local search structure for each out-of-core cluster. This design keeps raw vectors and local indices non-replicated on SSD, while representing cross-partition connectivity logically in memory (Section~\ref{sec:independent-partitions}).
\rev{\projectname{} also supports incremental updates through partition-local mutable deltas and graph-abstraction refreshes; local compaction periodically incorporates delta data into the corresponding SSD partition without rebuilding or repartitioning unrelated partitions (Section~\ref{sec:dynamic-update}).}

\noindent\textbf{Workflow.}
Given a query batch, \projectname{} first traverses the active memory-resident graph abstraction $\mathcal{GA}$ for each query and maps the reached probe nodes to clusters, producing a cluster-evidence list before SSD access.
The execution layer then schedules queries with similar predicted cluster visits, prunes low-value candidate clusters, and issues I/O requests only to the remaining selected partitions.
For each selected partition, \projectname{} invokes its scale-aware local search structure, such as Flat, graph, or auxiliary IVF, and merges local results into the global top-$k$ queue.
During serving, lightweight cluster-access statistics are recorded to update cluster hotness, refresh representative vectors on a shadow graph abstraction, and periodically publish the refreshed abstraction for future queries.

\subsection{The Memory-Resident Graph Abstraction}
\label{sec:dynamic-navigation}

GANNS maintains a compact memory-resident graph abstraction $\mathcal{GA}$ to provide logical connectivity across disjoint SSD partitions before local search. 
$\mathcal{GA}$ employs a navigable search graph over partition representatives: the graph enables efficient in-memory traversal, while the representatives determine which partition evidence is exposed before SSD access. 
GANNS, therefore, represents each cluster with a stable centroid anchor and selected representative vectors, which are connected by proximity edges and traversed entirely in memory before out-of-core local search. 
Given the disjoint cluster set $\mathcal{C}=\{C_i\}_{i=1}^{n}$ with centroids $\{c_i\}_{i=1}^{n}$, the node set of $\mathcal{GA}$ consists of all centroids and selected representative vectors:
\[
\mathcal{V}_{\mathcal{GA}}=
\{c_i\}_{i=1}^{n}\cup\bigcup_{i=1}^{n} R_i,\quad R_i\subseteq C_i .
\]
Here, each $c_i$ provides stable global coverage, while $R_i$ contains representative vectors selected from cluster $C_i$ by Algorithm~\ref{alg:representative} to capture off-center partition structure from diverse directions and radii.
During initialization, \projectname inserts all centroids into $\mathcal{GA}$ and selects the initial representative set $R_i$ for each cluster using Algorithm~\ref{alg:representative}. Each node stores the vector representation needed for in-memory distance computation, together with a
lightweight reference to its cluster ID and local vector ID. Thus, $\mathcal{GA}$ can be traversed entirely in memory, while raw vectors and local indices remain non-replicated on SSD.

\begin{algorithm}[t]
\caption{Representative Vector Selection for Graph Abstraction}
\label{alg:representative}
\KwIn{cluster $C_i$, centroid $c_i$, representative-vector budget $b$, candidate cap $M$, radius quantile $\rho$, score weights $\alpha,\beta$}
\KwOut{representative vector set $R_i$}
$P \leftarrow \textsc{DeterministicSample}(C_i, M)$\;
\ForEach{$x \in P$}{
    $r_x \leftarrow \|x-c_i\|_2$\;
    $u_x \leftarrow (x-c_i)/\max(r_x,\epsilon)$\;
}
$Q \leftarrow \{x \in P \mid r_x \ge \textsc{Quantile}(\{r_y:y\in P\}, \rho)\}$\;
$R_i \leftarrow \{\arg\max_{x\in Q} r_x\}$\;
\While{$|R_i| < b$ and $Q \setminus R_i \neq \emptyset$}{
    \ForEach{$x \in Q \setminus R_i$}{
        $D(x,R_i) \leftarrow 1-\max_{y\in R_i}\langle u_x,u_y\rangle$\;
        $\mathrm{score}(x) \leftarrow
        \alpha \cdot \textsc{Norm}(r_x)
        +\beta \cdot D(x,R_i)$\;
    }
    $R_i \leftarrow R_i \cup
    \{\arg\max_{x\in Q\setminus R_i}\mathrm{score}(x)\}$\;
}
\Return{$R_i$}\;
\end{algorithm}
Algorithm~\ref{alg:representative} selects representative vectors that complement the centroid of each cluster. 
It first samples a bounded candidate pool and computes each candidate's radius and normalized direction relative to the centroid. 
Since the centroid already represents the central tendency of the cluster, the algorithm uses the radius quantile $\rho$ to focus on off-center candidates that better capture cluster shape. 
It initializes $R_i$ with the farthest candidate and then greedily selects up to $b$ vectors using a score that combines two signals. 
The radius term favors candidates that extend coverage beyond the centroid, while the directional-diversity term prevents the selected vectors from collapsing into the same direction. 
The resulting set gives $\mathcal{GA}$ a compact, shape-aware summary of each physical partition.

\subsection{Out-of-Core Local Search Structures}
\label{sec:independent-partitions}
\begin{figure}[t]
    \centering
    \includegraphics[width=\linewidth]{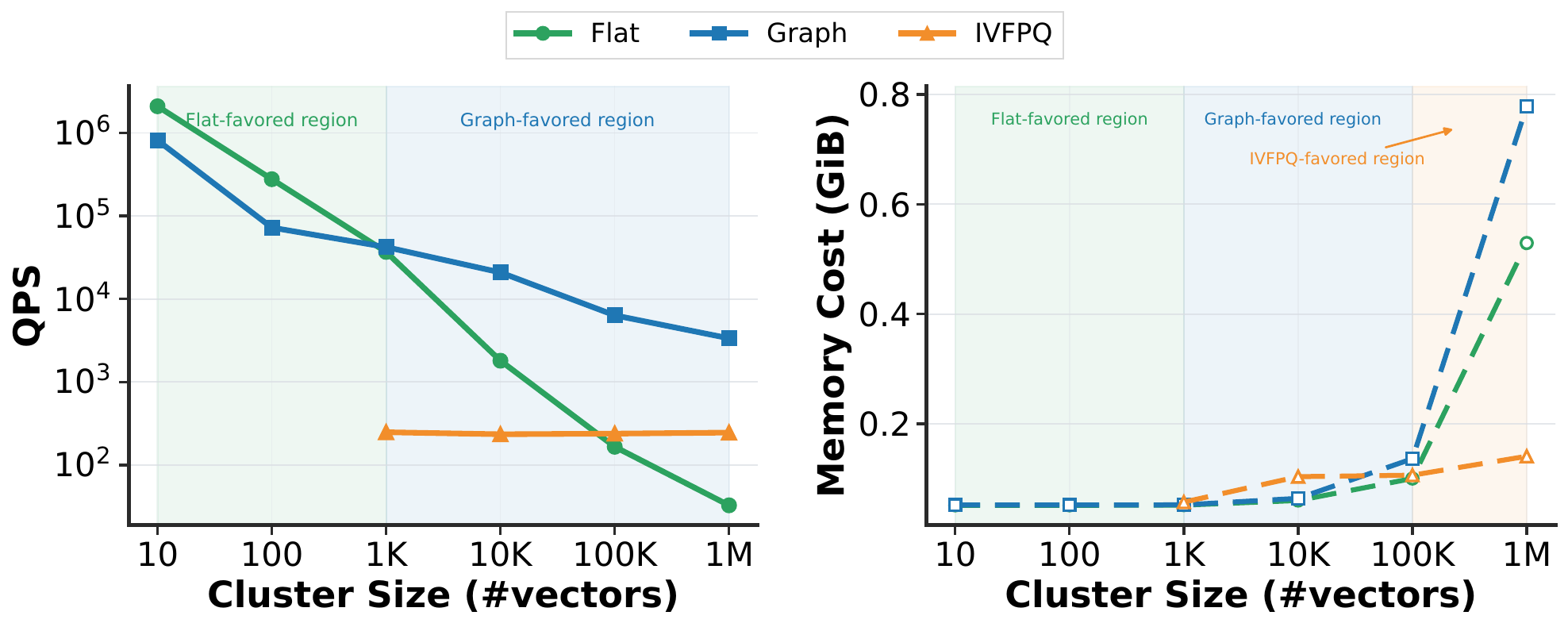}
    \caption{QPS and memory cost across index types vs. cluster size at Recall@100 $\geq$ 95\%.}
    \label{fig:scale}
\end{figure}

After the graph abstraction identifies the out-of-core partitions, \projectname has the flexibility to choose the local search structure
for each partition independently. As shown in Figure~\ref{fig:scale}, the best
structure changes with cluster scale under the same accuracy target: Flat scan is preferable for tiny clusters because it avoids index metadata and random traversal; graph search serves as the primary high-accuracy structure for most clusters; and a compact intra-cluster IVF view becomes useful for very large clusters when graph metadata
creates page-cache pressure. Therefore, \projectname assigns each cluster a scale-aware primary local structure and, for selected clusters, optionally builds a memory-efficient IVF view.

\noindent\textbf{Scale-aware primary local structure.} \projectname first assigns a primary local structure to every cluster. If the cluster size is below a threshold $\theta_{\mathrm{flat}}$,
\projectname uses a Flat structure that only stores the vector range and ID mapping. Query processing then performs a short sequential scan inside the cluster, avoiding unnecessary graph metadata for the long tail of tiny partitions. For all remaining clusters, \projectname uses a graph index by default. Graph search avoids exhaustive scans while preserving high recall, and the active graph working set remains small enough to benefit from page-cache reuse.

\noindent\textbf{Budget-aware auxiliary IVF view.}
For very large clusters, the graph index can become inefficient under a tight memory budget: even if graph search reduces distance computations, its metadata and neighbor traversal may touch many random pages and evict useful structures from the page cache. To handle this case, \projectname builds an additional memory-efficient auxiliary IVF view for the largest clusters.

Specifically, \projectname sorts clusters by size as
$|C_{(1)}|\ge |C_{(2)}|\ge \cdots \ge |C_{(n)}|$ and reserves a fraction $\rho_{\mathrm{auxiliary}}$ of the serving memory budget
$B_{\mathrm{serv}}$ for auxiliary-IVF directories. It then selects the largest prefix that fits this budget:
\[
m=\max\left\{t:\sum_{j=1}^{t}\widehat{M}_{\mathrm{auxiliary}}(C_{(j)})
\le \rho_{\mathrm{auxiliary}} B_{\mathrm{serv}}\right\}.
\]
\projectname then builds auxiliary IVF views for
$C_{(1)},\ldots,C_{(m)}$. Here,
$\widehat{M}_{\mathrm{auxiliary}}(C_i)$ estimates the resident memory cost
of the auxiliary IVF view for cluster $C_i$. This rule uses one tunable parameter, $\rho_{\mathrm{auxiliary}}$, while the actual number of auxiliary-indexed clusters is determined by the memory budget and the
cluster-size distribution.

The auxiliary IVF view does not replicate raw vectors. Instead, it stores a compact intra-cluster IVF directory whose posting lists reference the
same raw-vector range. During query execution, the graph index remains the default local structure. Under high memory pressure, or when a cluster is reached as a lower-priority candidate from the graph
abstraction, the execution layer can switch to the auxiliary IVF view to reduce resident metadata and make the access pattern closer to sequential list scans. Thus, the auxiliary IVF view complements graph search, providing a memory-efficient alternative when graph traversal would incur excessive out-of-core page misses under
constrained memory.





  

\subsection{Incremental Updates and Local Maintenance}
\label{sec:dynamic-update}

\rev{\noindent\textbf{Partition-local updates.}
Each SSD partition maintains a persistent local index together with a mutable delta for recent updates.
For an insertion, \projectname{} first traverses the graph abstraction to identify a small set of relevant partitions and assigns the new vector to the closest candidate partition, preserving single-partition ownership.
The vector is then added to that partition's delta.
Deletions are handled lazily through deletion marks for delta vectors and tombstones for persistent vectors, while vector updates are implemented as delete-plus-insert.
Thus, routine updates remain local to the affected partition and do not require immediate SSD index reconstruction.}

\noindent\textbf{\rev{Update-aware logical routing.}}
\rev{Recent updates can shift a partition's data distribution before they are incorporated into its persistent index.
\projectname{} therefore represents active delta data with lightweight routing proxies in the memory-resident graph abstraction.
\projectname{} uses the graph-abstraction refresh mechanism in Section~\ref{sec:query} to update this routing state as the data distribution evolves.
Queries use the refreshed graph evidence to select relevant partitions and search both the persistent local index and its active delta, filtering deleted entries before merging the results.
This makes recent updates immediately searchable while keeping logical routing consistent with the evolving data distribution.}

\noindent\textbf{\rev{Localized compaction.}}
\rev{When a partition's delta reaches a maintenance threshold, \projectname{} compacts only that partition by merging its active delta with the persistent data and rebuilding the corresponding local index on SSD.
Deleted and replaced records are removed during this process, and only the routing metadata of the affected partition needs to be refreshed.
The new local index and routing state are published together after construction completes, allowing the compacted delta to be retired without repartitioning or rebuilding the global index.
Thus, compaction remains confined to the affected partition and does not require maintenance of unrelated partitions.}

\section{System Optimization}
\label{sec:system-optimization}

The previous section described \projectname{}'s structure. We now show how this structure enables query-time optimizations by exploiting a workload property: cluster accesses are highly skewed, and their utility varies across clusters.

\noindent\textbf{Key observation: workload skewness.}
\begin{figure}[t]
    \centering
    \vspace{-0.8em}
    \begin{subfigure}[t]{0.49\linewidth}
        \centering
        \includegraphics[width=\linewidth]{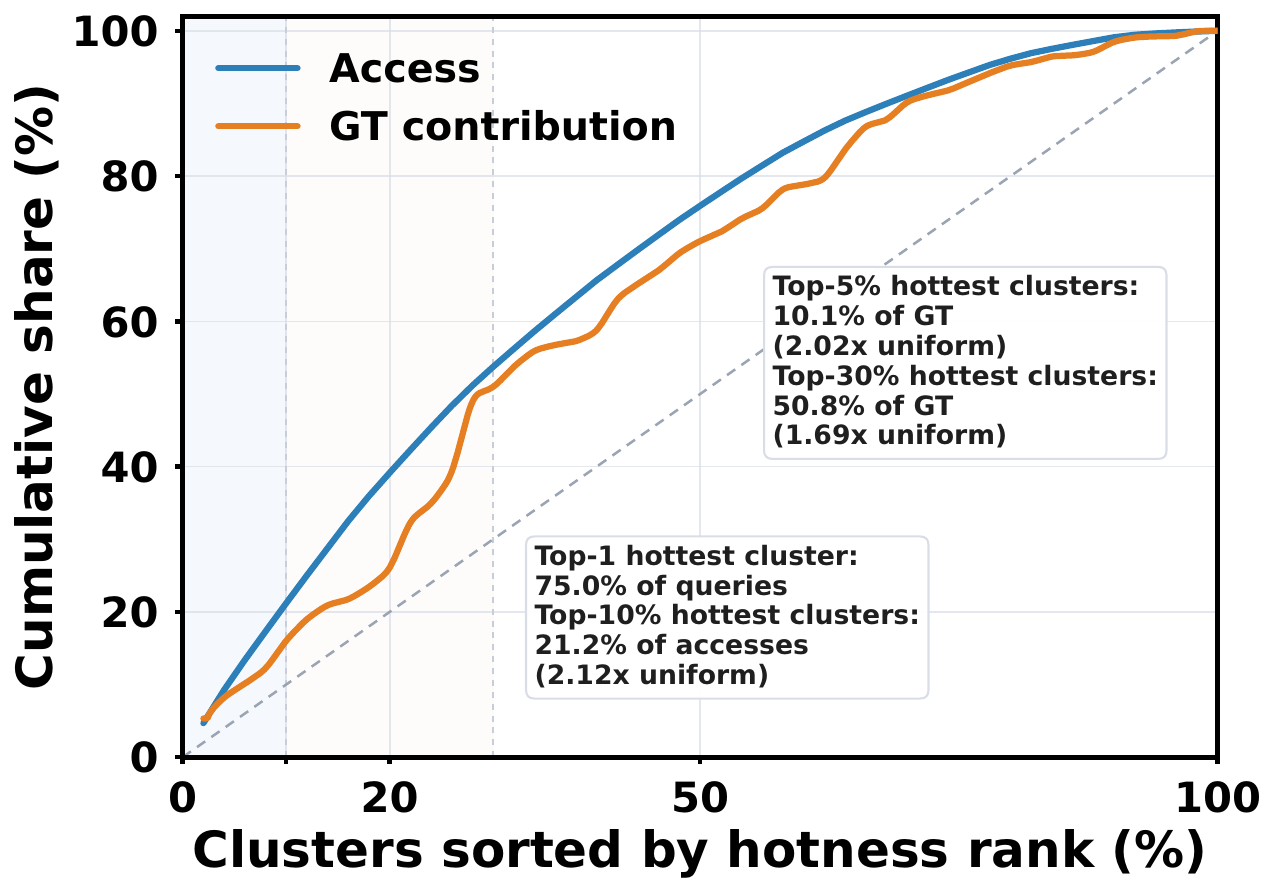}
        \caption{Access and GT concentration.}
        \label{fig:hot-cluster}
    \end{subfigure}
    \hfill
    \begin{subfigure}[t]{0.49\linewidth}
        \centering
        \includegraphics[width=\linewidth]{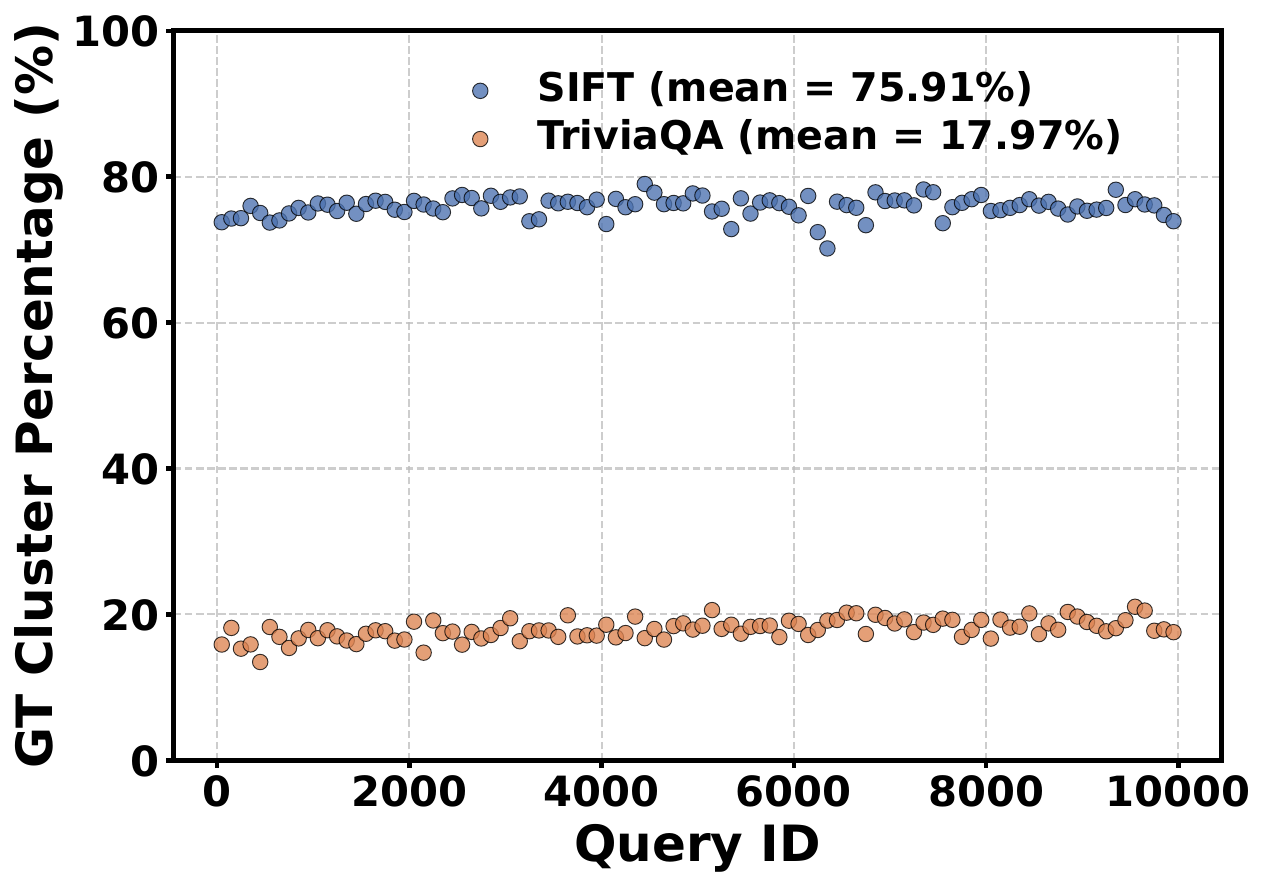}
        \caption{GT clusters Percentage.}
        \label{fig:gt-clusters}
    \end{subfigure}
    \vspace{-0.8em}
    \caption{Runtime access hotness and cluster utility. 
    A small set of clusters receives a large fraction of accesses and contributes disproportionately to ground-truth results, while many probed clusters, especially on semantic workloads, do not contribute to the final top-$k$ results.}
    \label{fig:access-hotness}
    \vspace{-0.8em}
\end{figure}
Figure~\ref{fig:hot-cluster} shows that query accesses are highly concentrated across clusters. On HotpotQA, the hottest cluster appears in $75.0\%$ of queries, and the top-$10\%$ hottest clusters receive $21.2\%$ of all accesses, which is $2.12\times$ higher than a uniform distribution.
More importantly, hotness is correlated with search utility rather than being only a cache-locality signal: the top-$5\%$ hottest clusters contribute $10.1\%$ of ground-truth results, and the top-$30\%$ hottest clusters contribute $50.8\%$ of ground-truth results. Thus, hot clusters are not merely frequently accessed; they are also more likely to contain useful neighbors.

Figure~\ref{fig:gt-clusters} further shows that routing alone is
insufficient. Even after the routing stage produces candidate clusters, many probed clusters contain no ground-truth neighbors. On SIFT, about $24\%$ of probed clusters do not contribute ground-truth results. On TriviaQA, the average ground-truth-cluster percentage drops to $17.97\%$, meaning that more than $80\%$ of probed clusters do not contribute final top-$k$ results. If these clusters are searched blindly, they still trigger out-of-core local-index accesses and SSD reads.

Together, these observations show that query hotness is both skewed and utility-aware. \projectname therefore exploits cluster-level runtime signals in three ways. First, it incrementally adapts the graph abstraction with representative vectors from frequently useful clusters (Section~\ref{sec:query}). Second, it prioritizes promising clusters and prunes low-value accesses during online search (Section~\ref{sec:cluster-pruning}). Third, it schedules batched queries with similar cluster-access patterns close to each other to improve temporal locality in the page cache and local index state (Section~\ref{sec:query-scheduling}). These optimizations reduce redundant SSD accesses and improve locality without changing the partition layout.


\subsection{Query-Aware Graph Abstraction Adaptation}
\label{sec:query}

\rev{\projectname{} allocates additional representatives according to
cluster-level query hotness, since frequently accessed clusters are also
more likely to contain useful neighbors, while every cluster retains its
centroid anchor and remains routable.}
During serving, each worker records how often a cluster is selected by graph routing or searched by the local index in the current batch. At the end of batch $t$, \projectname aggregates these counters and updates the
hotness of cluster $C_i$ as
\[
h_i^{(t+1)}=\lambda h_i^{(t)}+(1-\lambda)\cdot a_i^{(t)},
\]
where $a_i^{(t)}$ is the batch-level access count and $\lambda$ controls history decay. Every $\Delta_B$ batches, \projectname incrementally reallocates representative vectors based on cluster hotness: hot clusters receive new representatives, while representatives from cold clusters are retired. This update is lightweight because each insertion is performed entirely in memory and is comparable to an HNSW search over $\mathcal{GA}$; retired vectors are removed lazily by marking them inactive. To avoid blocking serving queries, the update is performed by a background thread on a shadow copy of $\mathcal{GA}$ and published through an atomic pointer swap; queries therefore traverse immutable snapshots. Since only a small fraction of vectors are adjusted in each round, the maintenance cost is amortized under concurrent serving. Figure~\ref{fig:workflow} illustrates this process: compared with a static graph abstraction, the query-aware design reallocates representatives from cold clusters to hot clusters. This replacement policy keeps $\mathcal{GA}$ bounded while shifting representative capacity toward frequently accessed clusters.

\subsection{Cluster Pruning Guided by Graph Abstraction}
\label{sec:cluster-pruning}

\begin{figure}[t]
    \centering
    \includegraphics[width=\columnwidth]{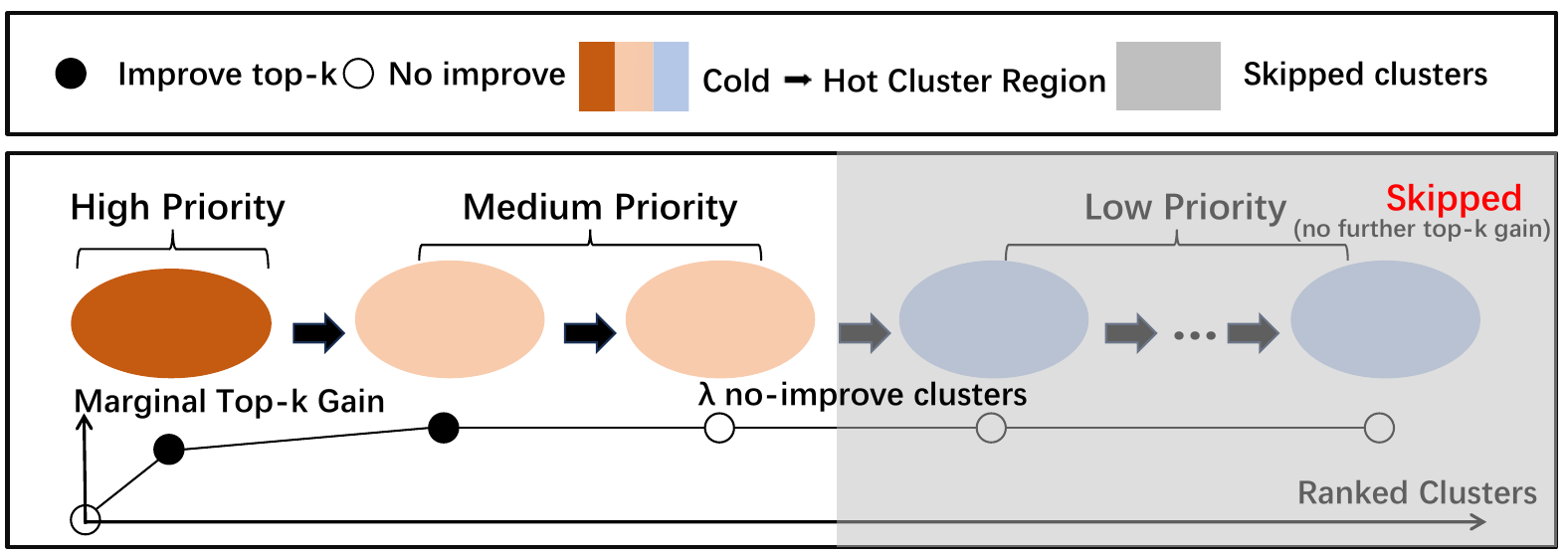}
    \caption{Graph abstraction guided cluster pruning.}
    \label{fig:cluster-prune}
\end{figure}
During query processing, \projectname traverses the in-memory graph
abstraction and obtains a set of probe vectors (Figure~\ref{fig:workflow}, step~2). Since each probe vector is mapped to a physical cluster, these probes provide lightweight cluster-level distance proxies before any out-of-core local search is invoked. For a cluster $C_i$, let $\mathcal{P}_i(q)$ denote the set of probe vectors reached by query $q$ and mapped to $C_i$. \projectname defines the cluster distance by the closest probe in this set:
\[
    D_i(q) = \min_{x \in \mathcal{P}_i(q)} \mathrm{dist}(q, x).
\]
Clusters are then visited in ascending order of $D_i(q)$
(Figure~\ref{fig:workflow}, step~3), so a cluster is searched earlier when at least one of its graph-abstraction representatives is close to the query. This avoids ranking clusters by $|\mathcal{P}_i(q)|$, which can bias the search toward clusters with more probes but weaker geometric relevance. The pruning policy is independent of the local index choice, and therefore applies uniformly to Flat, graph, and auxiliary-IVF structures.

\rev{After reordering, \projectname{} applies an early stopping rule to the low priority tail, using consecutive top-$k$ non-improvement as a signal of diminishing utility among the remaining candidates.}
Let $M$ be the number of candidate clusters and let $n=f(M)$, for example $n=\lceil \rho M \rceil$ for a fixed $\rho$. During cluster processing, \projectname maintains the global top-$k$ queue. If the next $n$ consecutively processed clusters do not improve the global top-$k$ queue, \projectname terminates cluster processing and skips the remaining candidates, as illustrated in Figure~\ref{fig:cluster-prune}.

As shown in Section~\ref{sec:component-ablation}, detailed ablation experiments confirm the effectiveness of this pruning strategy: many clusters with large proxy distances are never loaded from SSD, reducing unnecessary out-of-core searches.

\subsection{Query-Locality Scheduling}
\label{sec:query-scheduling}

Query access skewness causes many queries to repeatedly visit a small set of hot clusters. If queries are processed strictly in arrival order, visits to the same cluster may be separated by unrelated queries, forcing the system to reload similar local-index metadata and vector pages from SSD. \projectname therefore applies batch-level query-locality scheduling, which reorders nearby queries so that queries likely to access the same cluster are executed close together. This improves temporal locality in the page cache and local index state without changing any per-query search procedure.

Given a scheduling window of $W$ queries, \projectname first obtains a lightweight cluster visit plan for each query using the memory-resident graph abstraction $\mathcal{GA}$. This planning step runs in parallel and reuses the routing information already needed by normal query execution. 
For each query $q$, \projectname{} extracts the first cluster in its ordered cluster list, denoted as $\mathrm{top1}(q)$.
\rev{The cluster ranked first serves as a lightweight locality signature, as it is already exposed by routing and indicates the SSD region most likely to be accessed first.}
\projectname{} then stably sorts queries within the window by $\mathrm{top1}(q)$.
Queries with the same key preserve their original arrival order. The window size $W$ bounds scheduling delay and prevents global reordering, making the optimization suitable for high-throughput serving.

After scheduling, \projectname executes each query normally with the same search parameters, pruning rule, and local index procedure. The scheduler only changes the order of independent queries. By placing queries with similar cluster visits closer together, \projectname improves reuse of recently accessed local-index metadata and vector pages, thereby reducing redundant SSD traffic. This optimization is complementary to cluster pruning: pruning removes low-value cluster accesses, while query-locality scheduling improves locality among the clusters that are still searched.

\section{Evaluation}
\label{sec:exp}

\textbf{Implementation details.}
We implement \projectname as a C++17 CPU--SSD out-of-core search engine built with CMake. 
The codebase contains roughly 12K lines of project code, excluding third-party and generated code. 
\projectname uses FAISS~\cite{johnson2019faiss} for $k$-means clustering and auxiliary IVF-view indexes, HNSWLib~\cite{malkov2020hnsw} for the graph abstraction, and NSG~\cite{DBLP:journals/pvldb/FuXWC19} for local graph indexes inside clusters. 
HNSWLib is a natural fit for the graph abstraction because it provides efficient in-memory graph search and supports incremental insertion; stale abstraction nodes are invalidated lazily during refresh. 
Cluster vectors, id mappings, and local index artifacts are stored as disk files and accessed through memory mapping at query time. 
Unless otherwise stated, we use $\alpha=2.0$, $\beta=1.0$, and $\rho=0.7$ in Algorithm~\ref{alg:representative}. 
For auxiliary-IVF construction, we tune the memory-budget fraction $\rho_{\mathrm{aux}}$ empirically; under the default setting this typically builds auxiliary IVF views for about 5\%--10\% of clusters. 
For cluster pruning, we use an early-stop window $n=\lceil\rho_{\mathrm{prune}}M\rceil$ with $\rho_{\mathrm{prune}}=0.2$, where $M$ is the number of candidate clusters.

\noindent\textbf{Settings.}
All experiments are conducted on a server with two 64-core Intel Xeon Gold CPUs (2.5\,GHz), 512\,GB of DRAM, and a 1.92\,TB SATA SSD, running Ubuntu 20.04 LTS. 
Unless otherwise stated, all SSD-resident datasets and index artifacts are placed on this SSD for all methods. 
Following the cgroup-based memory-constrained methodology used in Starling~\cite{wang2024starling}, we run each method inside a Linux cgroup v2 memory controller with \texttt{memory.high}=4\,GB and \texttt{memory.max}=10\,GB by default. 
This setting applies to both anonymous memory and resident file-backed pages, so memory mapped vector and index pages are charged to the same budget after being faulted in. 
We monitor cgroup memory usage. 
For major-page-fault measurements, we read the \texttt{pgmajfault} field from the cgroup's \texttt{memory.stat} before and after each measurement run, and report per-query deltas. 
By default, query serving uses 48 worker threads.

\begin{table}[t]
\centering
\caption{Characteristics of datasets.}
\small
\begin{tabular}{lcccc}
\toprule
\textbf{Dataset}  & \textbf{Vector} & \textbf{Dimension} & \textbf{Size} & \textbf{Data Type} \\ \midrule
HotpotQA & 1.3M & 384 & 2 GB & Float \\  
SIFT & 100M & 128 & 12 GB &  Uint8\\
TriviaQA & 8.6M & 768 & 25 GB & Float \\
DEEP & 100M & 96 & 36 GB & Float \\
SPACEV & 1B & 100 & 372 GB & Float \\ \bottomrule
\end{tabular}
\label{table:datasets}
\vspace{-1.0em}
\end{table}

\noindent\textbf{Datasets.}
We evaluate \projectname on five public datasets ranging from millions to billions of vectors, each with 10K queries: HotpotQA~\cite{yang2018hotpotqa}, SIFT~\cite{jegou2011productSIFT}, TriviaQA~\cite{triviaqa2017}, DEEP~\cite{covington2016deep}, and SpaceV1B (SPACEV)~\cite{chen2018sptag} (Table~\ref{table:datasets}).
Across all datasets, IVF partitioning exhibits clear cluster-size skewness, while QA corpora such as HotpotQA and TriviaQA show particularly strong semantic skewness in addition to a pronounced long-tail distribution. We use a tighter 1 GB memory limit for HotpotQA.
For SPACEV, we convert the original \texttt{int8} vectors to \texttt{float} to align with common graph-based indexing libraries and ensure a fair comparison.

\noindent\textbf{ANNS Baselines.}
We compare \projectname against state-of-the-art out-of-core ANNS systems:

\begin{itemize}[leftmargin=*,topsep=0pt,itemsep=0pt,parsep=0pt]
    \item \textbf{DiskANN}~\cite{jayaram2019diskann}: A foundational SSD-based ANNS system and one of the first to support billion-scale vector search.
    \item \textbf{Starling}~\cite{wang2024starling}: An enhanced variant of DiskANN with improved indexing and navigation strategies.
    \item \textbf{SPANN}~\cite{chen2021spann}: An IVF-based ANNS system that trades disk space for performance by storing multiple replicas of the dataset to improve accuracy and latency.
\end{itemize}

For all baselines, we use the best or default configurations recommended by their open-source implementations, and sweep only the standard
search-strength parameter to obtain different recall points~\cite{jayaram2019diskann, wang2024starling}. For DiskANN and Starling, we fix the graph-search beam width to 4 and sweep the candidate-pool length $L$. 
\rev{We enable the default in-memory PQ representation for DiskANN and
Starling, but disable disk-side PQ compression because it substantially
degrades recall in our experiments.}
All methods are run on the same machine as \projectname.
We also include \textbf{PipeANN}~\cite{guo2025achieving}, an out-of-core ANNS engine that overlaps SSD I/O and computation through pipelined execution.

\begin{figure*}[!t]
  \centering
  \begin{subfigure}[t]{0.24\linewidth}
    \centering
    \includegraphics[width=\linewidth]{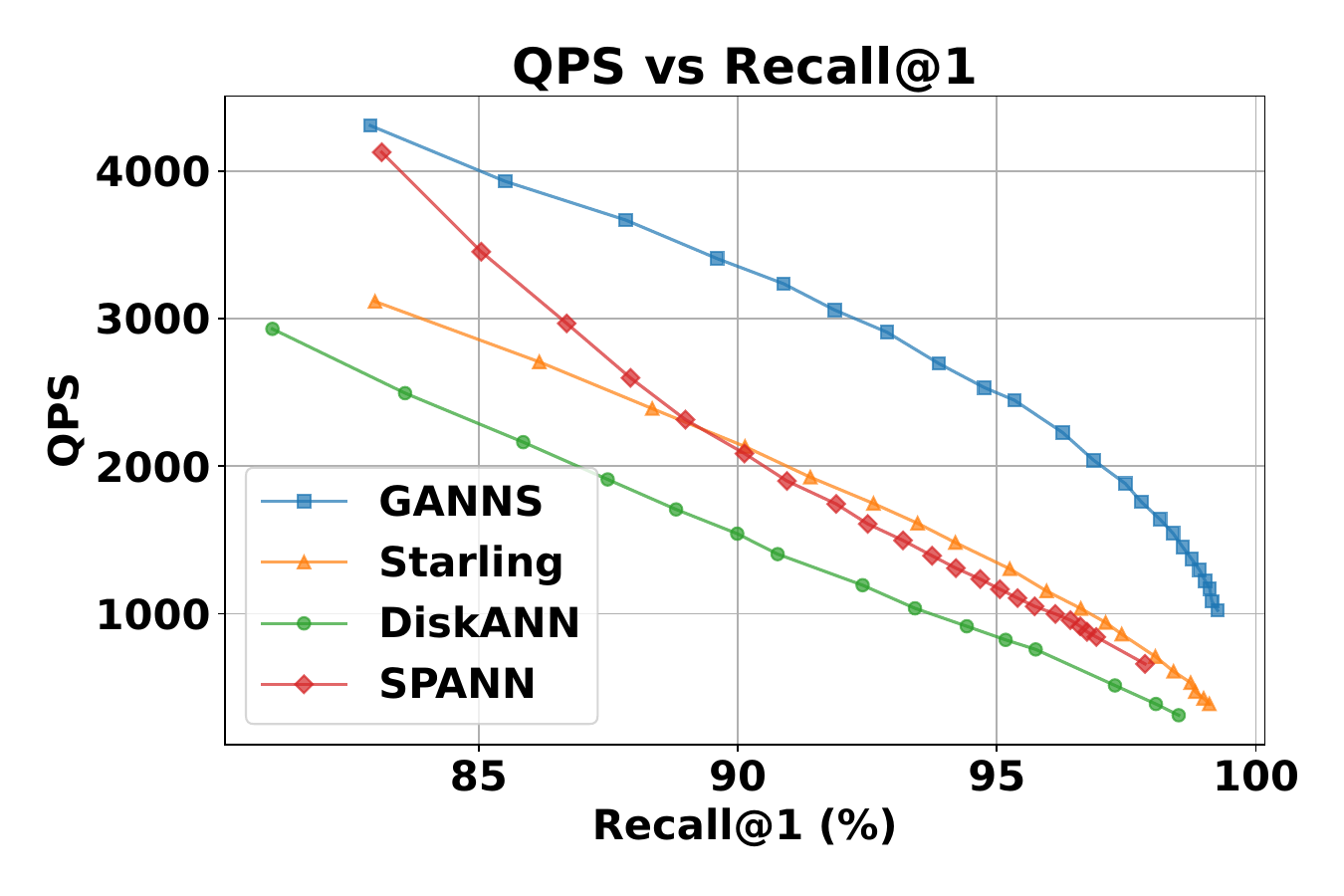}
    \caption{SIFT: QPS-Recall.}
    \label{fig:bigann_qps_k1}
  \end{subfigure}%
  \hfill
  \begin{subfigure}[t]{0.24\linewidth}
    \centering
    \includegraphics[width=\linewidth]{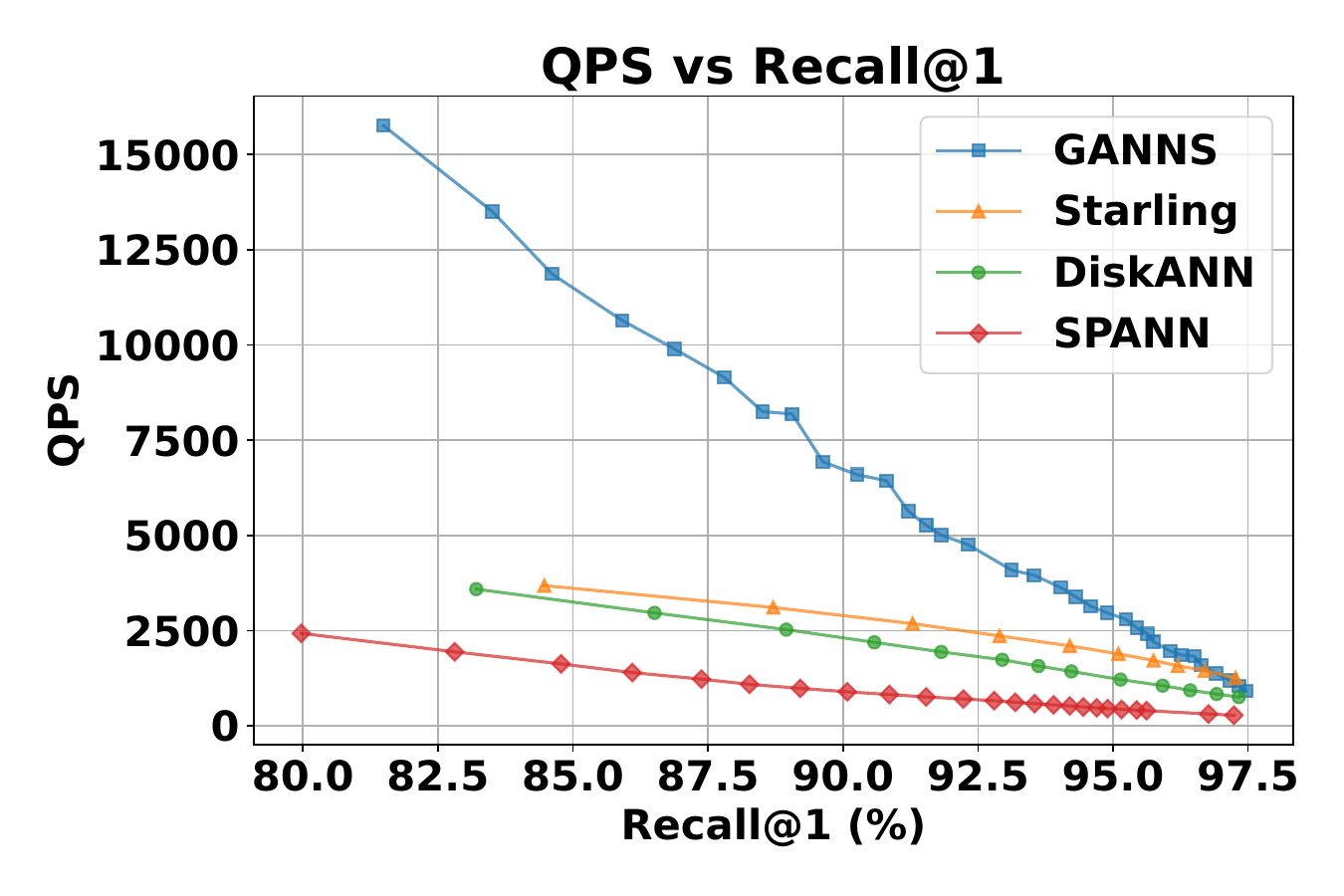}
    \caption{DEEP: QPS-Recall.}
    \label{fig:deep_qps_k1}
  \end{subfigure}%
  \hfill
  \begin{subfigure}[t]{0.24\linewidth}
    \centering
    \includegraphics[width=\linewidth]{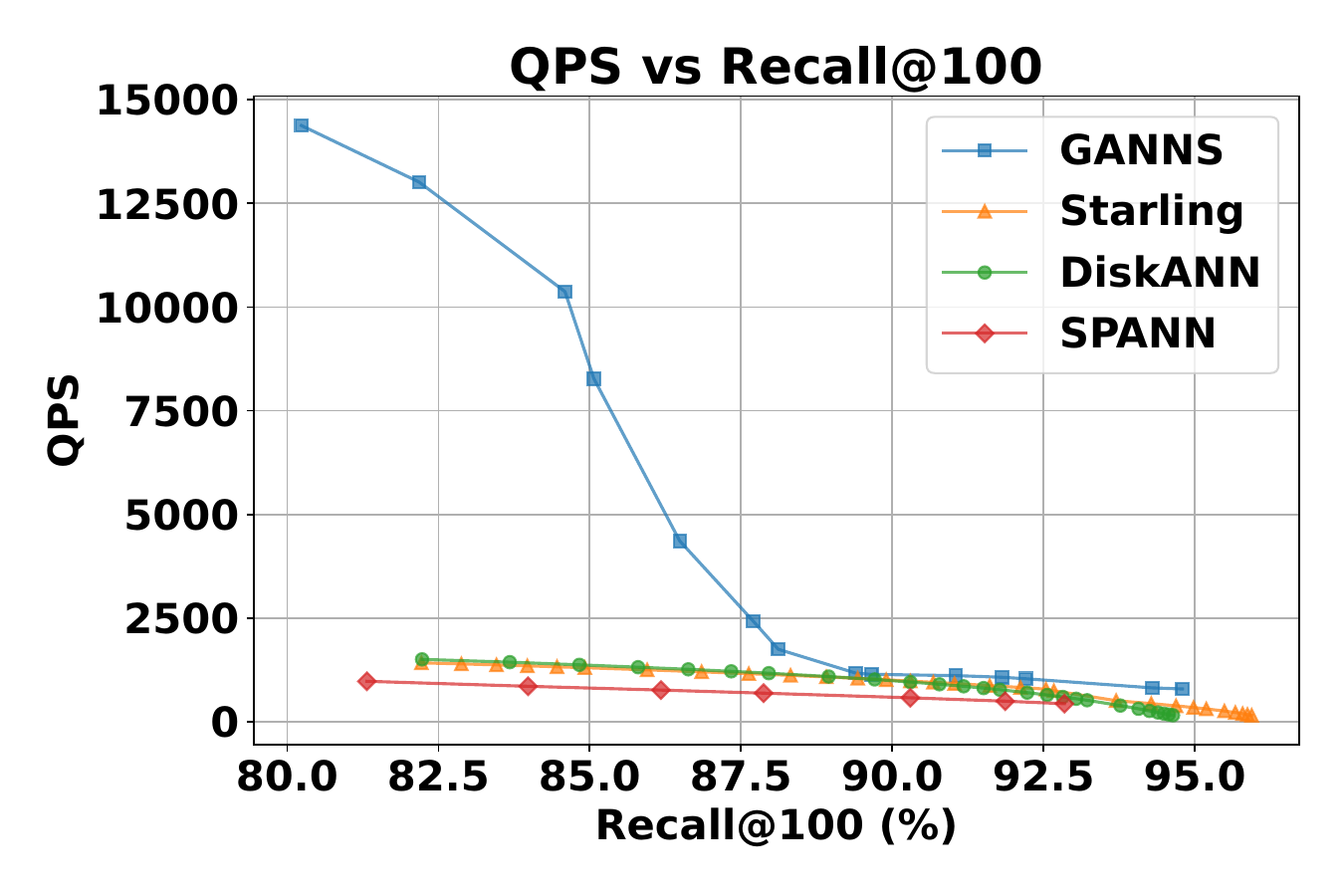}
    \caption{TriviaQA: QPS-Recall.}
    \label{fig:triviaqa_qps_k100}
  \end{subfigure}
  \begin{subfigure}[t]{0.24\linewidth}
    \centering
    \includegraphics[width=\linewidth]{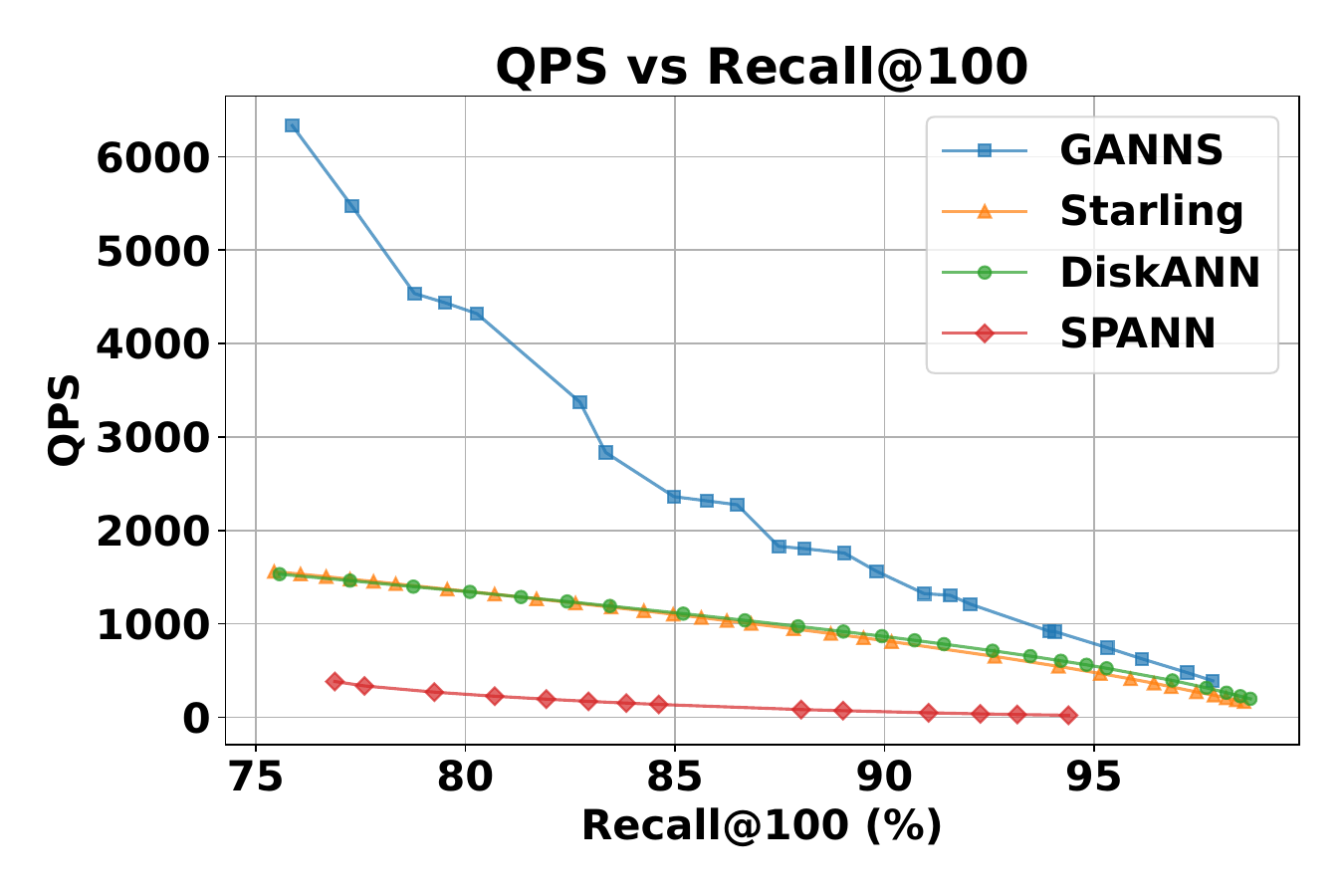}
    \caption{HotpotQA: QPS-Recall.}
    \label{fig:hotpotqa_qps_k100}
  \end{subfigure}
  \caption{QPS overall evaluation on SIFT, DEEP, TriviaQA and HotpotQA datasets.}
  \label{fig:QPS-overall}
\end{figure*}

\begin{figure*}[!t]
  \centering
  \begin{subfigure}[t]{0.24\linewidth}
    \centering
    \includegraphics[width=\linewidth]{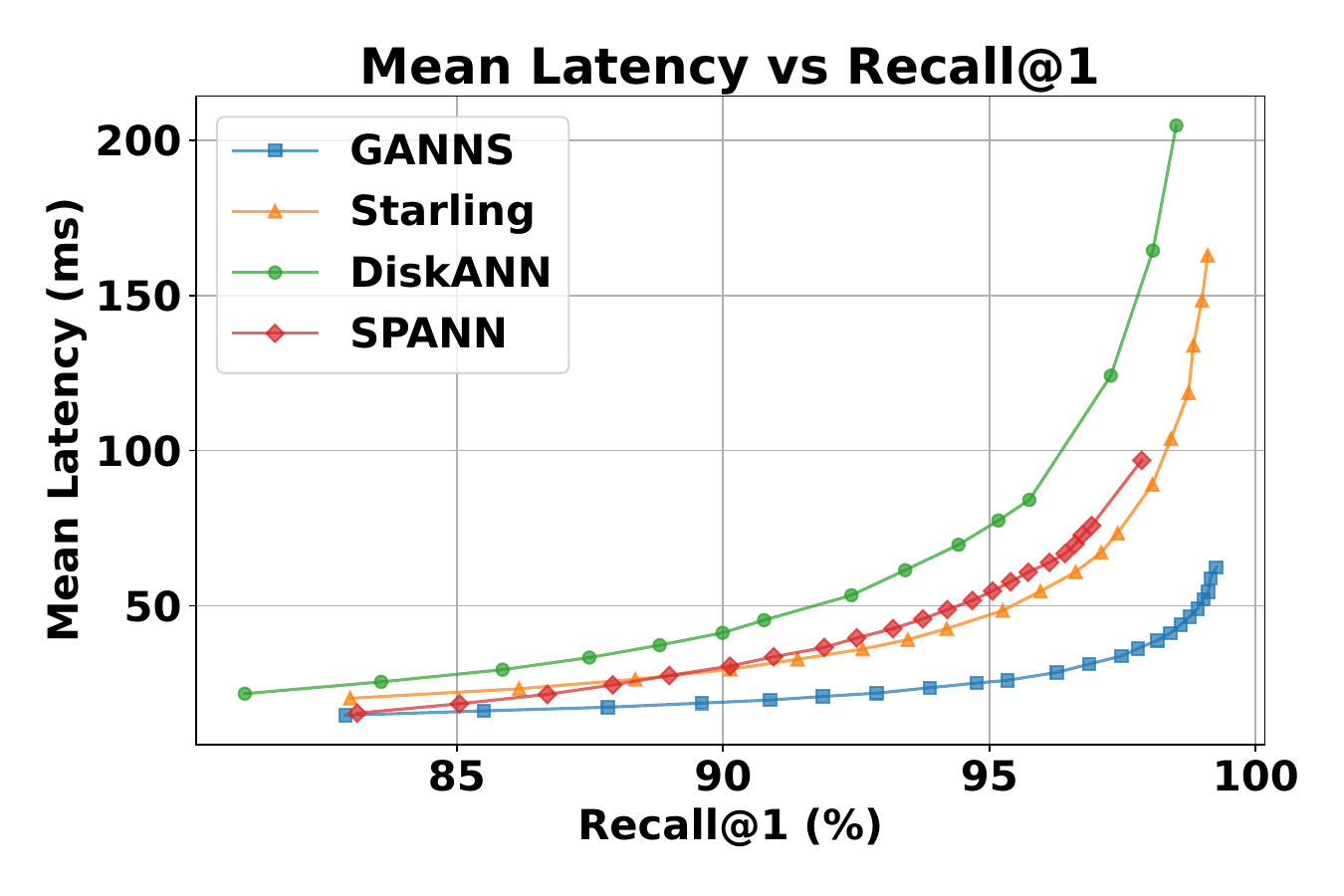}
    \caption{SIFT: Latency-Recall.}
    \label{fig:bigann_latency_k1}
  \end{subfigure}%
  \hfill
  \begin{subfigure}[t]{0.24\linewidth}
    \centering
    \includegraphics[width=\linewidth]{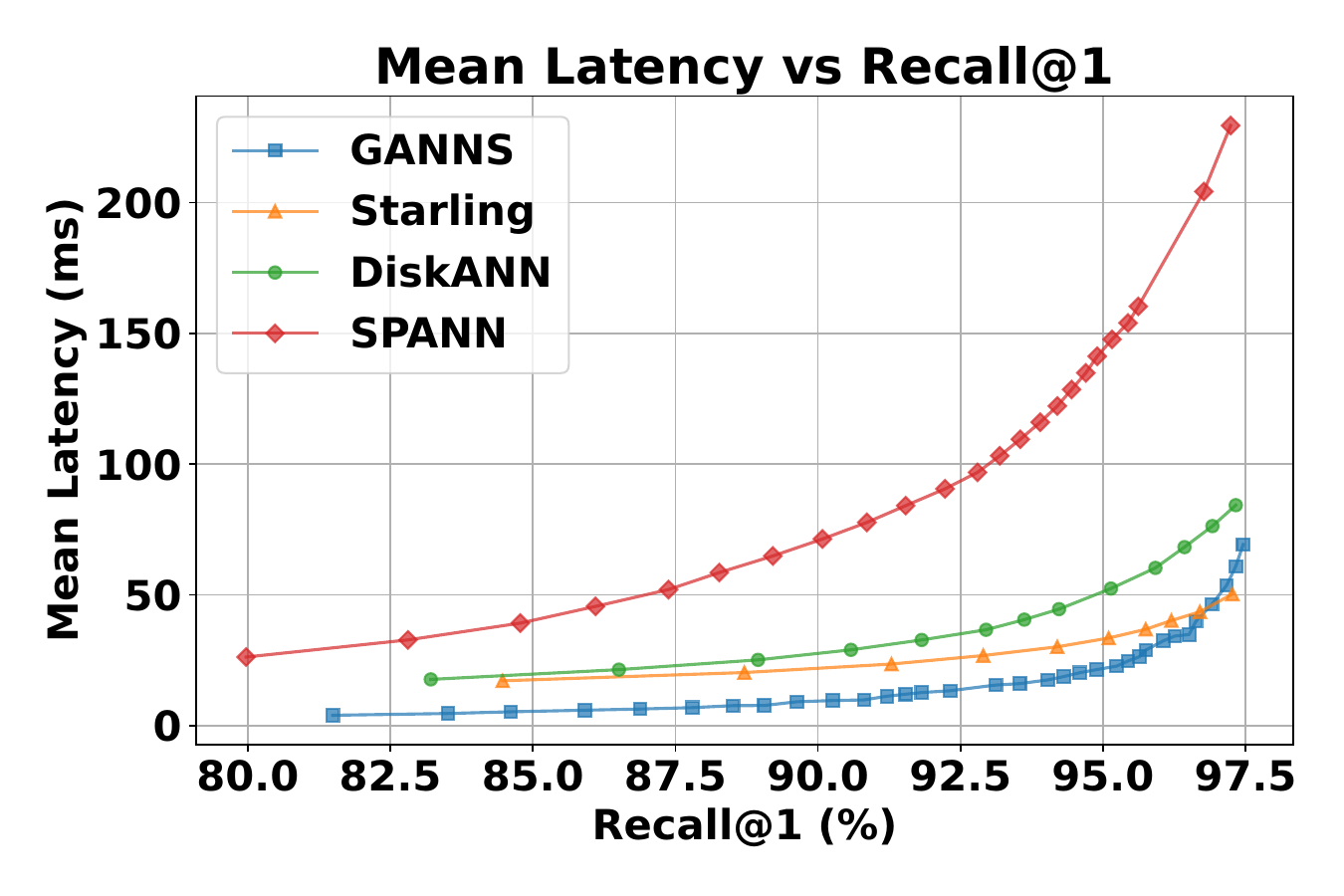}
    \caption{DEEP: Latency-Recall.}
    \label{fig:deep_latency_k1}
  \end{subfigure}%
  \hfill
  \begin{subfigure}[t]{0.24\linewidth}
    \centering
    \includegraphics[width=\linewidth]{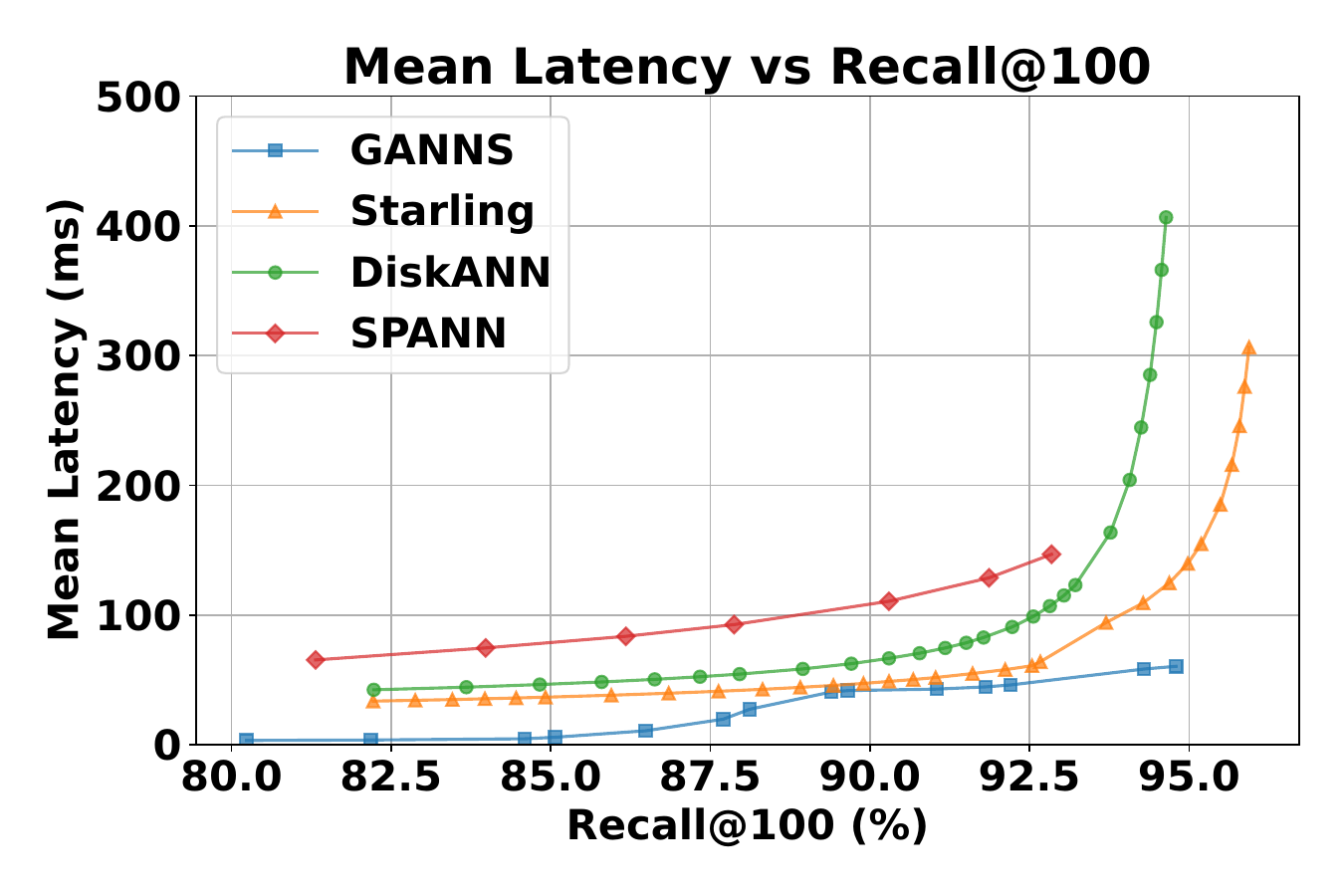}
    \caption{TriviaQA: Latency-Recall.}
    \label{fig:triviaqa_latency_k100}
  \end{subfigure}
  \begin{subfigure}[t]{0.24\linewidth}
    \centering
    \includegraphics[width=\linewidth]{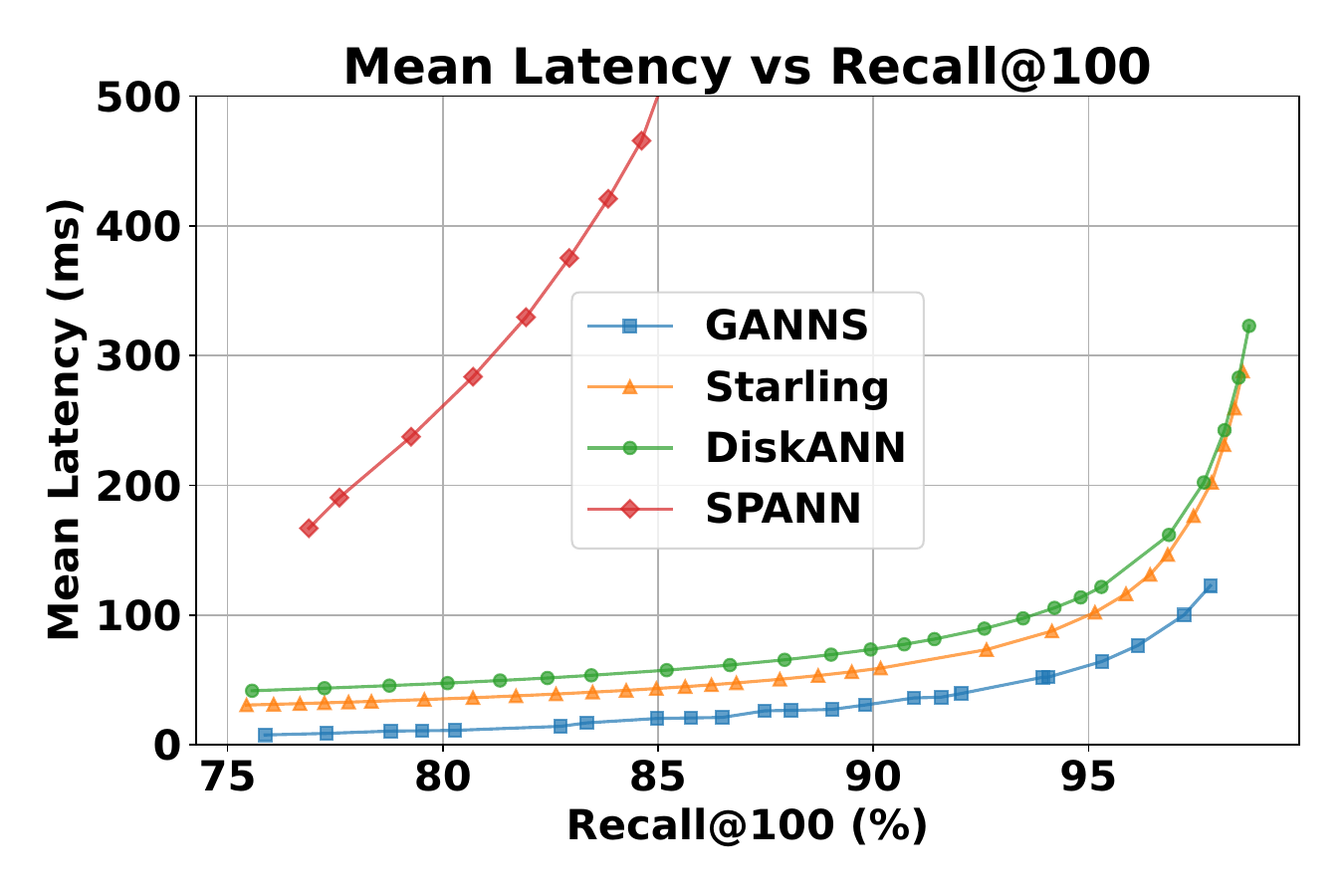}
    \caption{HotpotQA: Latency-Recall.}
    \label{fig:hotpotqa_latency_k100}
  \end{subfigure}
  \caption{Latency overall evaluation on SIFT, DEEP, TriviaQA and HotpotQA datasets.}
  \label{fig:Latency-overall}
\end{figure*}

\begin{figure}[t]
  \centering
  \vspace{-1em}
  \begin{subfigure}[t]{0.49\linewidth}
    \centering
    \includegraphics[width=\linewidth]{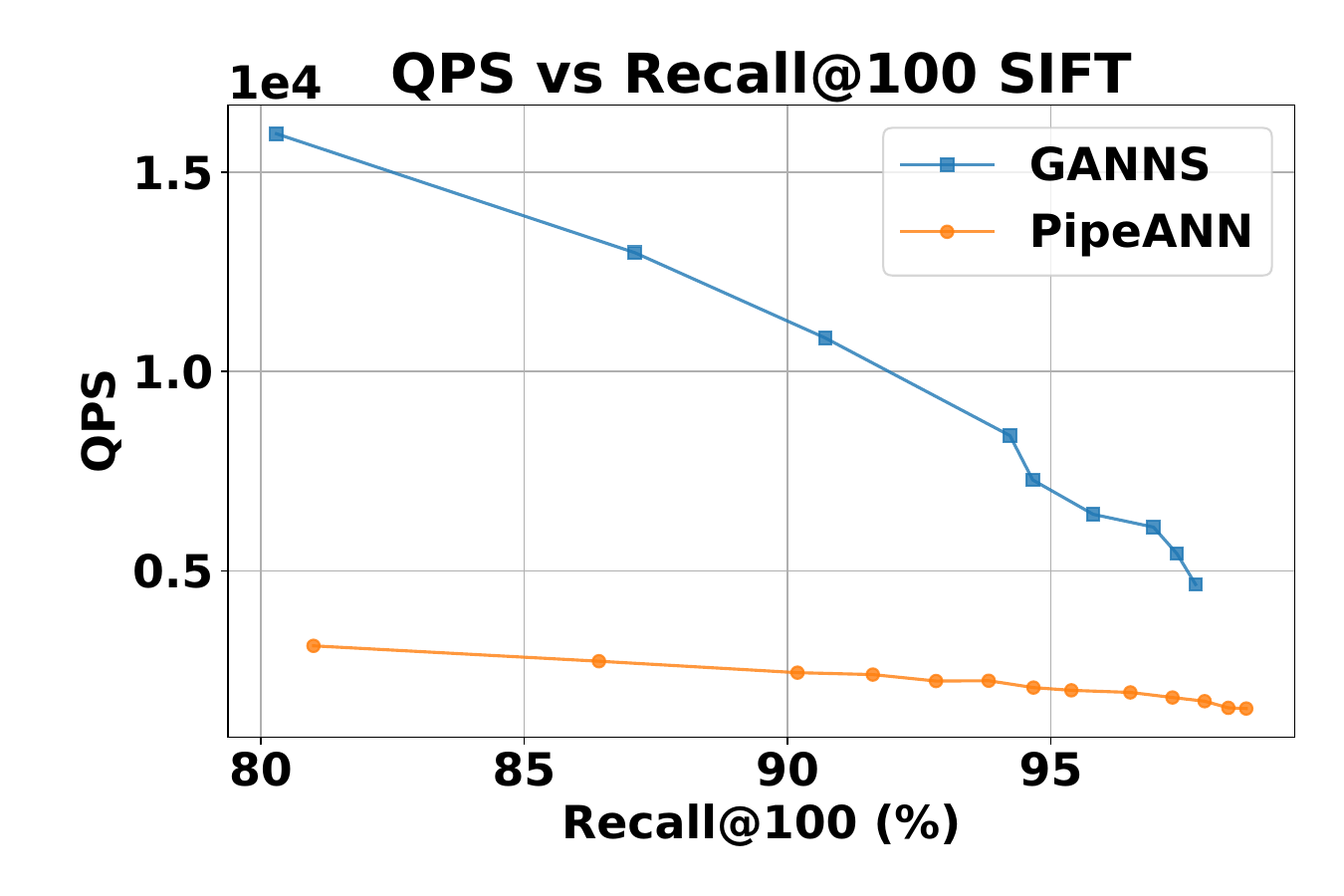}
    \caption{SIFT: QPS-Recall.}
    \label{fig:sift_pipeann}
  \end{subfigure}%
  \hfill
  \begin{subfigure}[t]{0.49\linewidth}
    \centering
    \includegraphics[width=\linewidth]{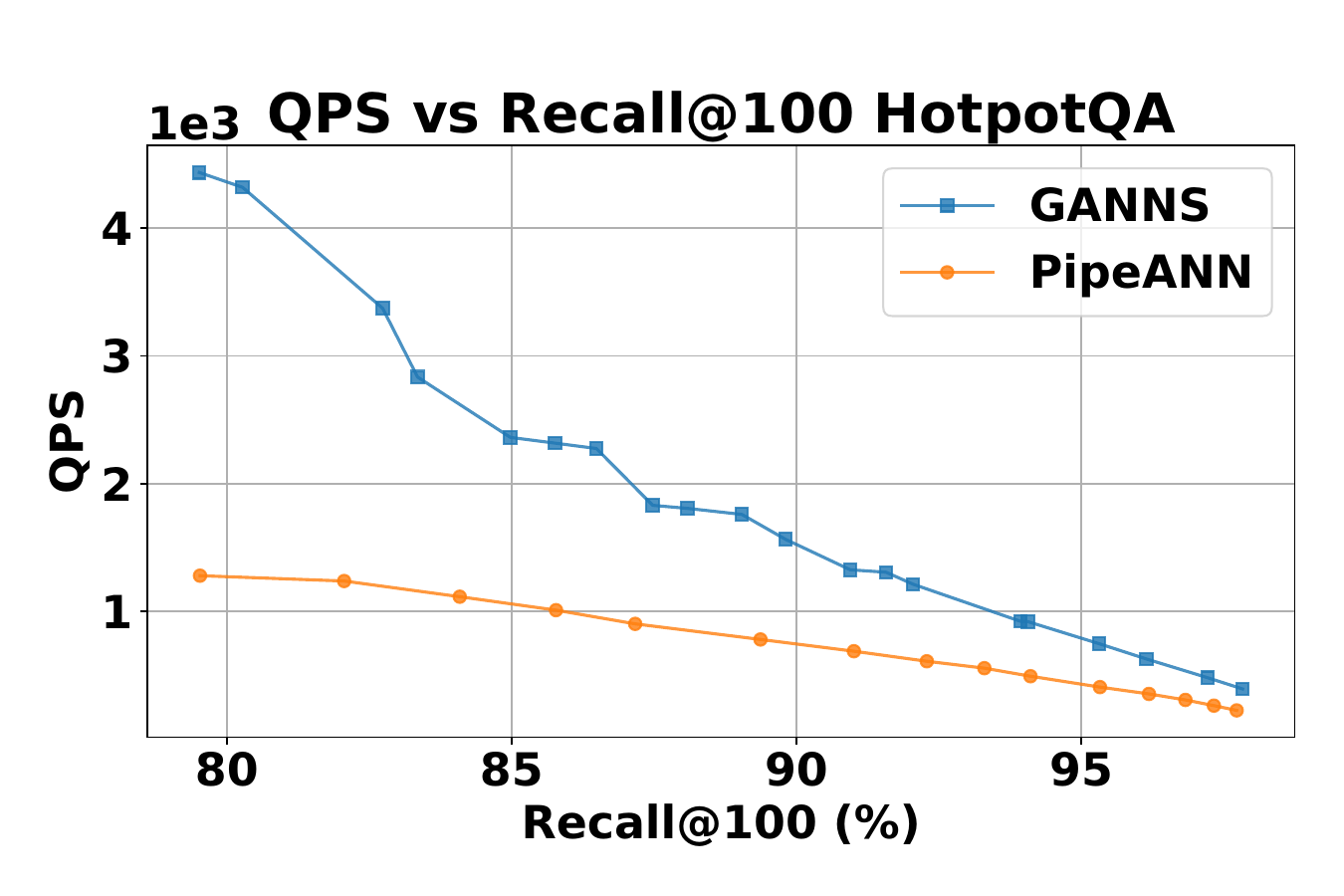}
    \caption{HotpotQA: QPS-Recall.}
    \label{fig:hotpotQA_pipeann}
  \end{subfigure}%
  \caption{QPS evaluation comparing with PipeANN.}
   \vspace{-1em}
  \label{fig:compare_pipeann}
\end{figure}

\subsection{Comparison With SOTA}

As shown in Figures~\ref{fig:QPS-overall} and~\ref{fig:Latency-overall}, we compare \projectname{} against Starling~\cite{wang2024starling}, DiskANN~\cite{jayaram2019diskann}, and SPANN~\cite{chen2021spann} on five datasets: SIFT, DEEP, SPACEV, TriviaQA, and HotpotQA.
\projectname{} consistently achieves higher QPS and lower latency across recall targets.
On feature datasets such as SIFT and DEEP, \projectname{} delivers \textbf{2.0$\times$--4.7$\times$} higher QPS than Starling/DiskANN and reduces latency by up to \textbf{12.3$\times$} compared with SPANN.
The gains are larger on semantic workloads: on TriviaQA and HotpotQA, \projectname{} achieves up to \textbf{17.2$\times$} higher QPS and \textbf{25.0$\times$} lower latency than SPANN at Recall@100 $=0.85$.

These improvements come from reducing unnecessary SSD work at multiple levels.
Compared with SPANN, \projectname{} avoids replicated posting-list accesses by keeping raw vectors in non-replicated SSD partitions and providing boundary reachability through memory-resident logical overlap.
Compared with DiskANN and Starling, \projectname{} avoids relying only on SSD-resident graph traversal or static routing: its representative-vector graph abstraction produces stronger partition evidence, and the execution layer uses this evidence to prioritize partitions and prune low-utility accesses.
Finally, scale-aware local indexes avoid using one fixed structure for clusters with different sizes and memory footprints.

The benefit is most pronounced on HotpotQA and TriviaQA, where dense semantic regions and skewed partitions amplify the I/O cost of coarse routing and replication.
The advantage also extends to regular feature datasets: on SIFT, \projectname{} achieves up to \textbf{2.7$\times$} QPS gain, showing that the architecture improves robustness across both skewed semantic workloads and traditional feature benchmarks.

We also compare \projectname with PipeANN~\cite{guo2025achieving} on SIFT and HotpotQA, as shown in Figure~\ref{fig:compare_pipeann}. On SIFT, \projectname achieves up to \textbf{5.2$\times$} higher QPS at Recall@100 $=0.9$. On HotpotQA, it consistently delivers around \textbf{2$\times$} higher QPS across the evaluated recall range. 
\subsection{Ablation Study}
\label{sec:component-ablation}

\begin{table}
\centering
\scriptsize
\setlength{\tabcolsep}{3.5pt}
\caption{I/O behavior of architecture skeletons. P, G, and O denote Partition, Graph, and our partition-local index substrate, respectively; O is not the full \projectname{} pipeline.}
\label{tab:arch-io}
\begin{tabular}{llrrrrrr}
\toprule
Dataset & Mem. 
& \multicolumn{3}{c}{Miss/q} 
& \multicolumn{3}{c}{Miss rate} \\
\cmidrule(lr){3-5}\cmidrule(lr){6-8}
 & & P & G & O & P & G & O \\
\midrule
SIFT     & 90\% & 1009.40  & 498.42 & 38.41 & 45.23\% & 18.47\% & 3.15\%  \\
SIFT     & 50\% & 1865.57  & 1392.12 & 131.99 & 83.59\% & 51.58\% & 10.84\% \\
TriviaQA & 90\% & 6204.83  & 60.10 & 21.85 & 2.74\% & 2.79\% & 1.58\% \\
TriviaQA & 50\% & 31502.00 & 145.39 & 36.68 & 13.89\% & 6.76\% & 2.66\% \\
\bottomrule
\end{tabular}
\end{table}

\begin{figure*}[t]
  \centering
  \begin{subfigure}[t]{0.49\textwidth}
    \centering
    \includegraphics[width=\linewidth]{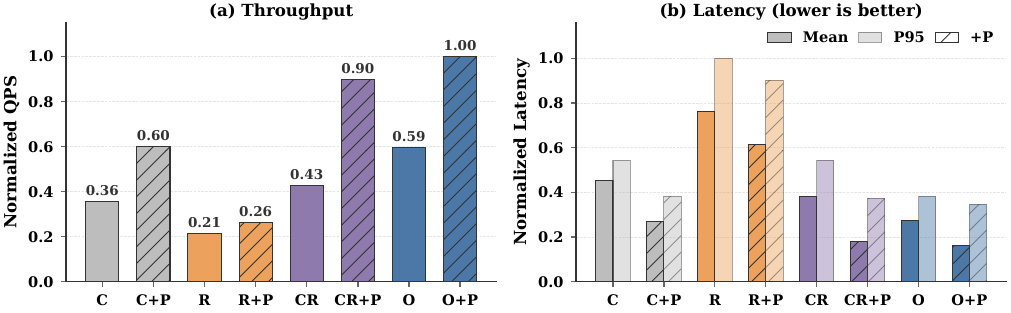}
    \caption{GA strategy and pruning.}
    \label{fig:ablation-ga-strategy}
  \end{subfigure}
  \hfill
  \begin{subfigure}[t]{0.49\textwidth}
    \centering
    \includegraphics[width=\linewidth]{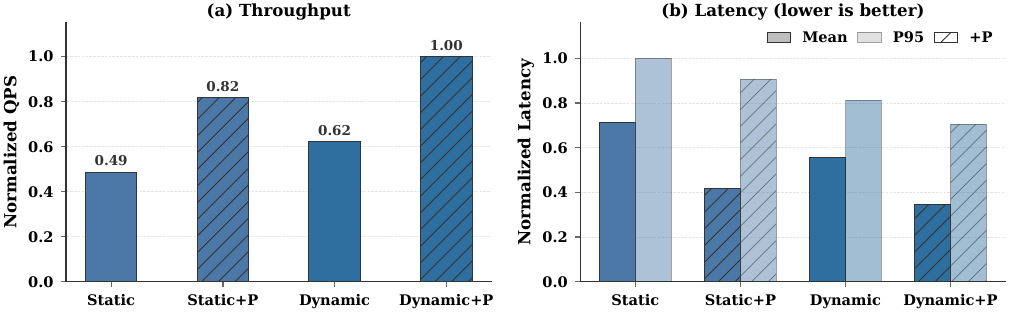}
    \caption{Static vs. dynamic GA.}
    \label{fig:ablation-ga-dynamic}
  \end{subfigure}
  \caption{Ablation of graph abstraction and pruning. C, R, CR, and O denote centroid-only, random-sample, centroid+random, and \projectname{}'s representative-vector graph abstraction, respectively; +P enables pruning.}
  \label{fig:ablation-upper}
\end{figure*}


\begin{figure}[t]
  \centering
  \includegraphics[width=\linewidth]{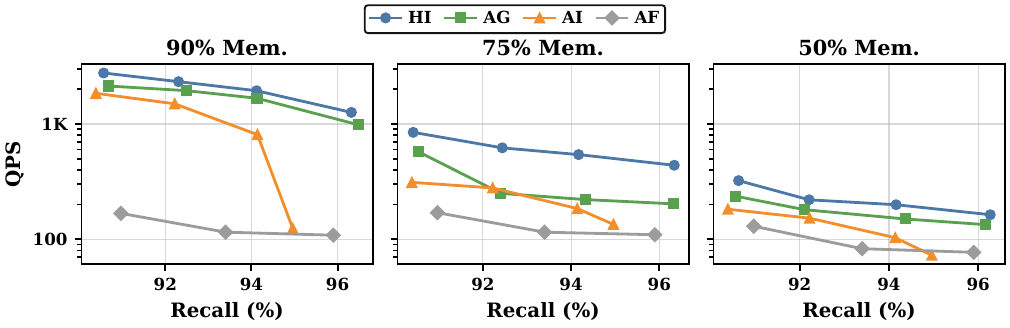}
  \caption{Ablation of the out-of-core partition layer under different memory budgets. HI denotes \projectname's hybrid indexing, while AG, AI, and AF denote all-graph, all-IVF, and all-Flat variants.}
  \label{fig:ablation-hybrid}
\end{figure}

\begin{figure}[t]
  \centering
  \includegraphics[width=\linewidth]{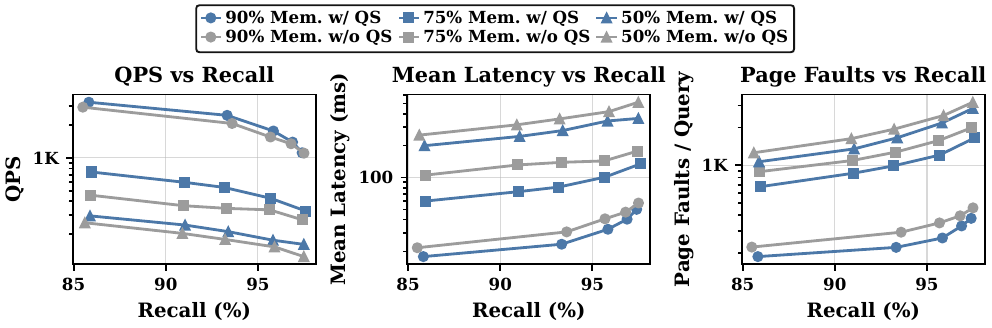}
  \caption{Ablation of query-locality scheduling under different memory budgets. QS reorders nearby queries with similar cluster visits to improve temporal locality without changing per-query search parameters.}
  \label{fig:ablation-query-scheduling}
\end{figure}

We first isolate the I/O behavior of the architectural substrate motivated in Section~\ref{sec:motivation-efficiency}, and then decompose \projectname{} into three parts that correspond to its hierarchical design: (1) the memory-resident graph abstraction and its cluster-level pruning support, (2) the out-of-core partition layer with hybrid local access paths, and (3) query-locality scheduling in the execution layer. 
Unless otherwise stated, component-level ablations are conducted on TriviaQA~\cite{triviaqa2017}. 

\noindent\textbf{Architectural substrate: out-of-core I/O pattern.}
Table~\ref{tab:arch-io} compares three pure architecture skeletons under different memory limits. 
Partition uses IVF-style partition search, Graph uses a monolithic HNSW-style graph layout, and Ours uses disjoint SSD partitions with local graph indexes inside selected partitions. 
This experiment isolates the physical substrate rather than the full \projectname{} pipeline.
\rev{Also, with pruning, scheduling, and other optimizations disabled, the partition-local graph substrate achieves 934.8 QPS on TriviaQA at around 90\% Recall@100, comparable to DiskANN's 1{,}020.6 QPS and 1.36$\times$ higher than SPANN's 689.5 QPS.}

The partition-local graph substrate consistently reduces unnecessary out-of-core accesses. 
On SIFT, Ours reduces misses/query from 1009.4 for Partition and 498.4 for Graph to 38.4 under the 90\% memory limit; under the 50\% memory limit, it reduces misses/query from 1865.6 and 1392.1 to 132.0. 
On TriviaQA, Ours also achieves the lowest miss rate under both memory limits, reducing misses/query to 21.9 and 36.7 under 90\% and 50\% memory, respectively. 
These results show that partition-local graph search preserves coarse partition locality while avoiding exhaustive partition scans and global random graph traversal. 
Therefore, disjoint SSD partitions with local indexes provide a suitable substrate for \projectname{}.

\noindent\textbf{Upper layer: graph abstraction and cluster pruning.}

\rev{Figure~\ref{fig:ablation-ga-strategy} compares GA strategies with and without pruning. Centroid+random improves over both centroid-only and random sampling alone, showing that the two signals are complementary; our radius-and-diversity selection improves further, indicating that representative choice matters beyond simply adding routing nodes.}
All variants are evaluated around the same 0.90 recall target.
Enabling pruning is nearly lossless: the recall drop is below 0.001, and the pruned variants still remain above 0.90 recall.
More generally, cluster pruning exposes an accuracy--efficiency tradeoff: even when a more aggressive policy becomes mildly lossy, the saved SSD budget can be reallocated to deeper local search or more candidate vectors in high-priority clusters instead of low-value partitions.
At this recall level, \projectname{} without pruning searches 16.0 clusters/query, $2.2\times$ fewer than random-sample routing; with pruning enabled, this drops to 8.1 clusters/query, $3.4\times$ fewer than random-sample routing with pruning.
These results show that better cluster evidence improves QPS and latency, and also makes pruning more effective by ranking useful clusters earlier before SSD access.
\rev{At Recall@100 near 90\%, representative routing, cluster ordering, and pruning together improve QPS by 49.9\% over centroid routing, reducing index reloads from 3{,}934 to 2{,}440.}

Figure~\ref{fig:ablation-ga-dynamic} separates the effect of dynamic updates. Dynamic GA alone improves over the static abstraction, and Dynamic+P performs best. 
This is because query-hot clusters are frequently visited and more likely to contribute useful results, so refreshing their representative vectors improves the ranking of candidate clusters before SSD access. 
We also measure cluster-selection precision and F1 score during query processing. 
Initial precision and F1 are about 0.16 and 0.24; after query-aware updates, they rise to about 0.20 and 0.30. 
This indicates that the graph abstraction becomes more accurate over time, reducing unnecessary cluster probes and strengthening the pruning signal without sacrificing recall.

\noindent\textbf{Out-of-core layer: hybrid local access paths.}
Figure~\ref{fig:ablation-hybrid} compares \projectname's hybrid indexing against uniform local-index choices under 90\%, 75\%, and 50\% memory budgets. Hybrid indexing consistently achieves the highest QPS across recall targets, and the advantage becomes more important as memory becomes tighter. All-Graph performs well when its working set can be reused, but it suffers under memory pressure because graph metadata and neighbor traversal touch many random pages. All-Flat avoids metadata but becomes inefficient once clusters grow beyond the tiny-partition regime. All-IVF provides a more compact access pattern, but using it uniformly sacrifices the accuracy-oriented behavior of graph search on medium clusters. \projectname avoids these extremes by using Flat for tiny clusters, Graph as the primary high-accuracy path, and memory-efficient auxiliary IVF views only for large clusters whose graph working set is likely to cause page-cache pressure.
\rev{In the repeated near matched recall comparison, converting the largest 5\% of clusters, which cover 8.59\% of the vectors, to auxiliary IVF views raises QPS by 27.0\%, cuts mean latency by 21.3\%, and reduces working set reloads from 2{,}598 to 1{,}714. The other 95\% retain graph search, targeting auxiliary indexing at clusters that dominate memory pressure.}

\noindent\textbf{Execution layer: query-locality scheduling.}
Figure~\ref{fig:ablation-query-scheduling} shows the effect of query-locality scheduling under multiple memory budgets. Enabling QS consistently improves QPS and reduces mean latency compared with processing queries in arrival order. The improvement is more visible under tighter memory budgets, where reusing recently touched local-index pages is more important. Since QS does not change per-query search parameters, the gain comes from better temporal locality among cluster accesses rather than from changing the search space. This optimization is complementary to cluster pruning: pruning reduces which clusters are accessed, while scheduling improves the order in which the remaining cluster accesses are issued.
\rev{Across three runs, a window of 1{,}024 queries raises QPS by 17.9\%, cuts mean execution latency by 20.2\%, and lowers working set reloads from 2{,}598 to 2{,}030, while searched clusters remain at 10.38 per query. Unchanged search work links the gain to reuse.}


\subsection{I/O Analysis}
\label{sec:eval-IO}

\begin{figure}[t]
  \centering
  \begin{subfigure}[t]{0.45\linewidth}
    \centering
    \hspace*{0.03\linewidth}
    \includegraphics[width=\linewidth]{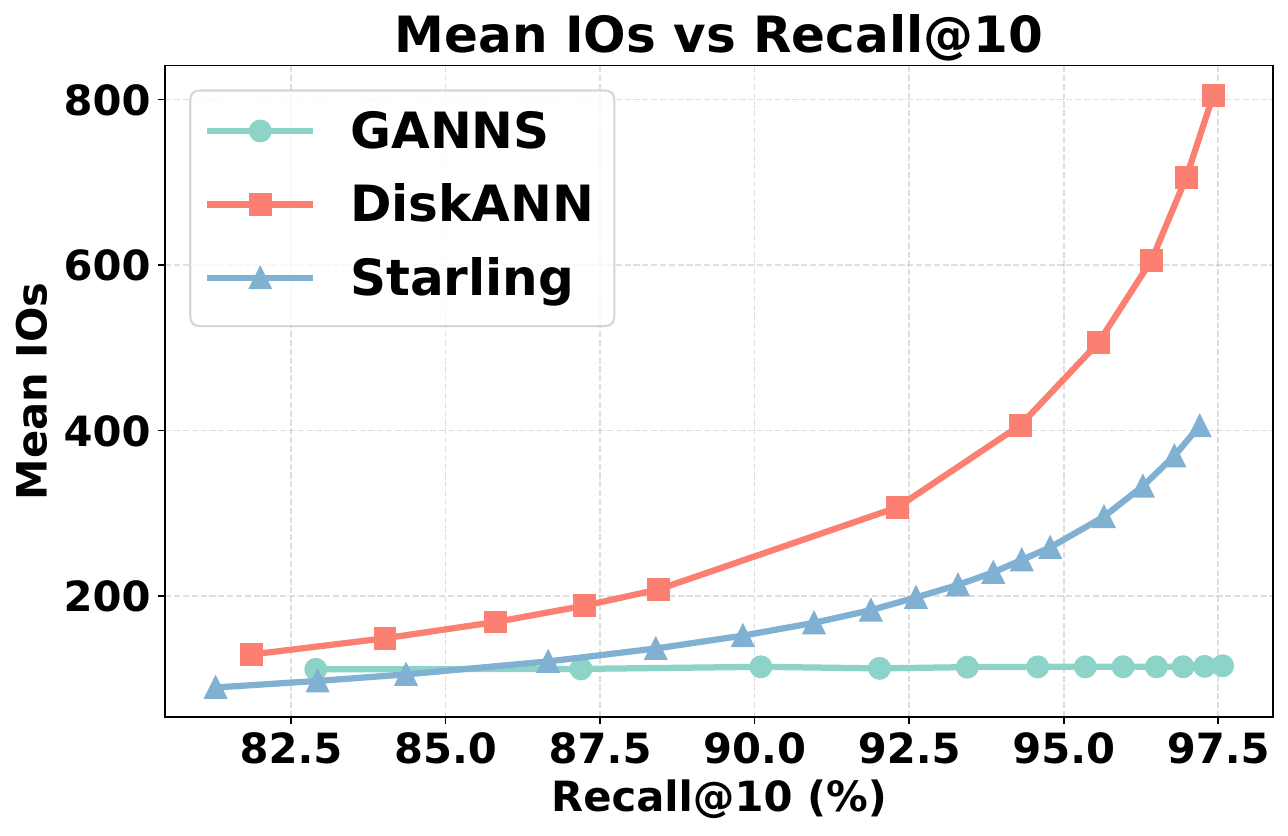}
    \caption{SIFT: IO-Recall.}
    \label{fig:bigann_mean_io_recall}
  \end{subfigure}
  \hfill
  \begin{subfigure}[t]{0.45\linewidth}
    \centering
    \includegraphics[width=\linewidth]{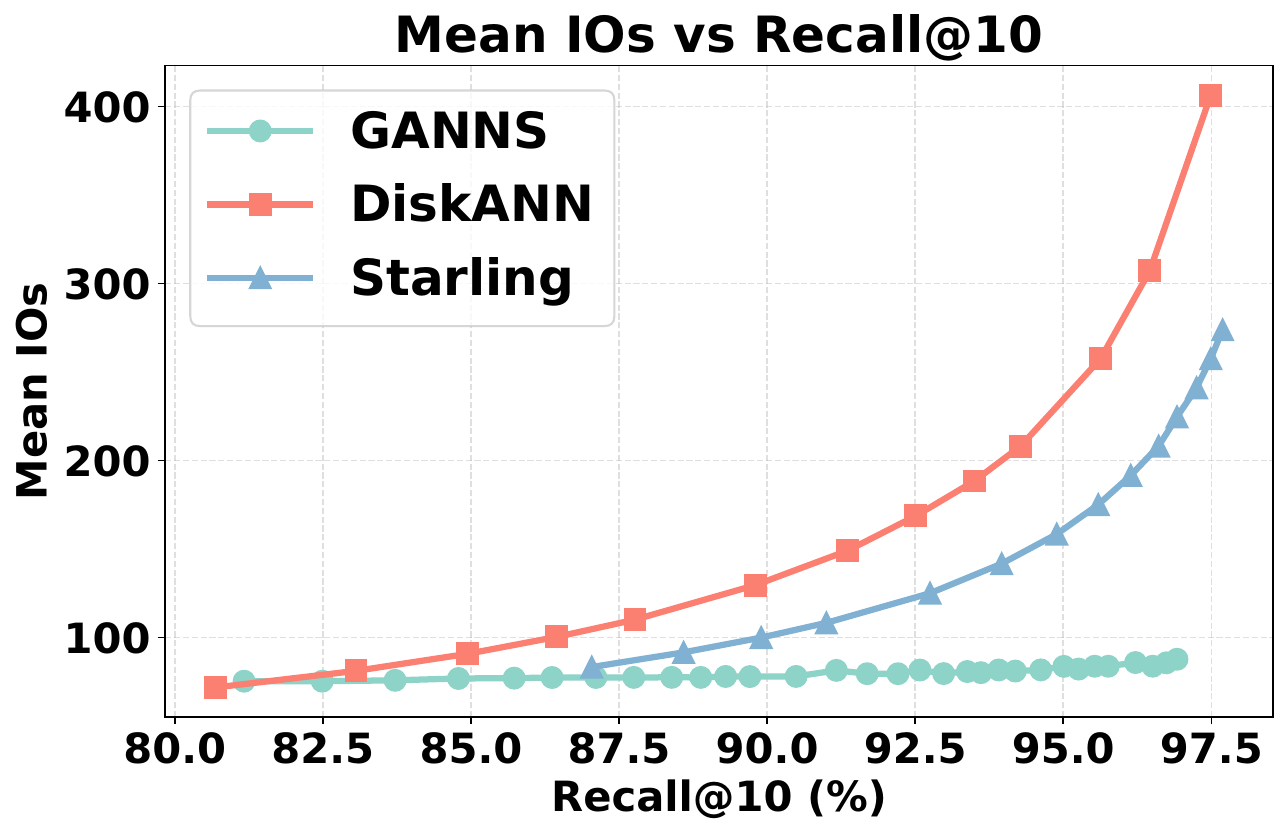}
    \caption{DEEP: IO-Recall.}
    \label{fig:deep_mean_io_recall}
  \end{subfigure}
  \caption{Page-access comparison on different datasets.}
  \vspace{-1em}
  \label{fig:mean_io_recall}
\end{figure}

We evaluate I/O efficiency by measuring the average number of page accesses per query for all systems. Figure~\ref{fig:bigann_mean_io_recall} and Figure~\ref{fig:deep_mean_io_recall} compare \projectname with DiskANN~\cite{jayaram2019diskann} and Starling~\cite{wang2024starling}. We omit SPANN~\cite{chen2021spann} from the plotted curves because its page-access counts are much larger and would obscure the other systems. On DEEP, SPANN already accesses 584 pages/query at Recall@10 $=0.928$, and this number grows to 964 and 1{,}905 pages/query at Recall@10 $=0.958$ and $0.982$, respectively; these values exceed the plotted range needed to compare the remaining systems.

At Recall@10 $=97\%$, \projectname achieves the same accuracy with fewer page accesses, requiring \textbf{3.5$\times$} fewer accesses than Starling and \textbf{7$\times$} fewer than DiskANN. The reduction comes from the hierarchical design: the memory-resident routing layer provides more accurate cluster hints before SSD access, while the out-of-core partition layer avoids a one-size-fits-all local index and lowers the cost of searching each cluster.

\projectname also keeps page accesses stable as recall increases. From Recall@10 $=90\%$ to $98\%$, its I/O grows by less than \textbf{10\%}. This stability indicates that higher recall does not require blindly probing many more disk-resident regions. Instead, logical cross-partition connectivity guides queries to relevant partitions, and cluster-level pruning skips low-utility candidates before issuing local searches. In contrast, DiskANN and Starling must touch increasingly more pages as recall targets rise, because their disk search paths or static routing signals provide weaker control over which out-of-core regions are actually useful.

\begin{figure}
  \centering
  \begin{subfigure}[t]{0.47\linewidth}
    \centering
    \includegraphics[width=\linewidth]{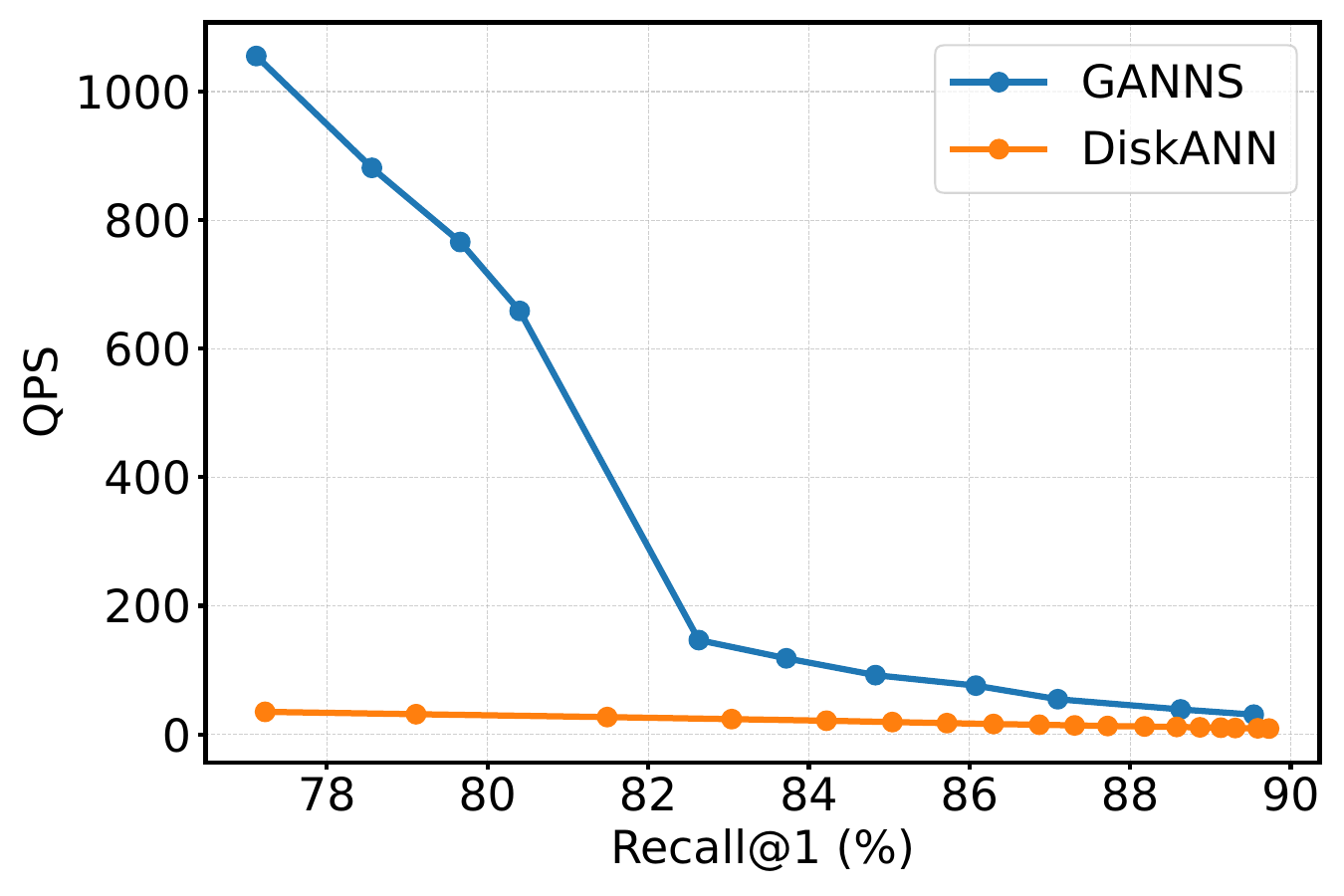}
    \caption{QPS vs Recall.}
    \label{fig:spacev_qps_recall}
  \end{subfigure}
  \begin{subfigure}[t]{0.47\linewidth}
    \centering
    \includegraphics[width=\linewidth]{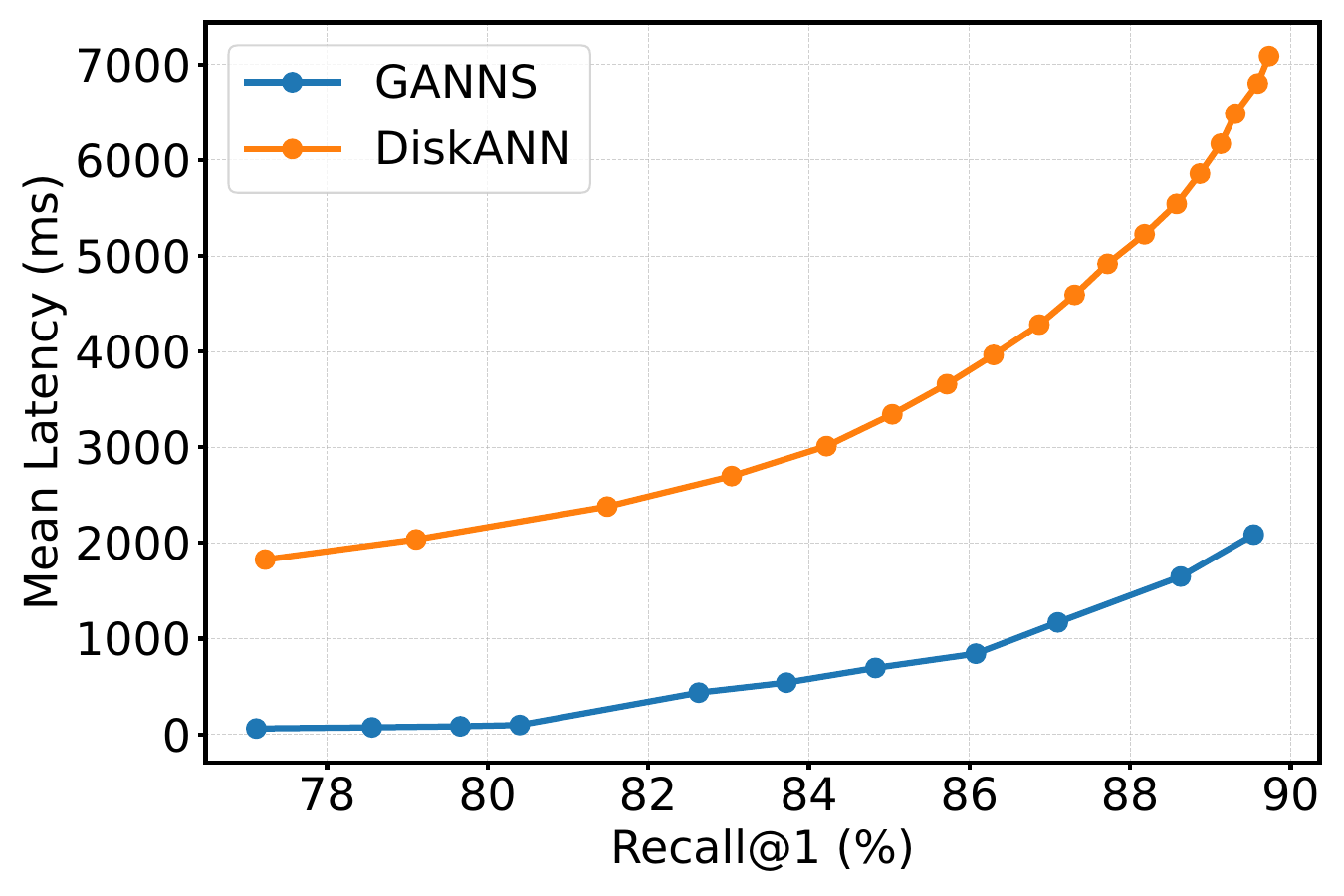}
    \caption{Latency vs Recall.}
    \label{fig:spacev_latency_recall}
  \end{subfigure}
  \caption{Performance evaluation on billion-scale data.}
  \label{fig:spacev_latency_scalability}
\end{figure}

\subsection{Evaluation on Billion-Scale Data}
Most existing ANNS systems, such as SPANN~\cite{chen2021spann}, require substantial memory during index construction, making them impractical for billion-scale datasets under limited hardware. For example, Starling~\cite{wang2024starling} failed to finish index construction on the SPACEV dataset within 120 hours. In contrast, \projectname is designed for resource-constrained environments and operates under a strict 10\,GB memory cap.

Figure~\ref{fig:spacev_qps_recall} and Figure~\ref{fig:spacev_latency_recall} report QPS and latency under these constraints. At Recall@1 $=0.8$, \projectname achieves a \textbf{24.1$\times$} higher QPS than DiskANN, and even at Recall@1 $=0.9$ it sustains a \textbf{5$\times$} advantage. Latency shows a similar trend: at comparable recall targets, \projectname delivers up to \textbf{3.5$\times$} lower mean latency. These results demonstrate that \projectname remains effective at a billion scale and under tight DRAM limits, allowing \projectname to maintain high efficiency where existing systems degrade sharply.

\begin{table}[t]
\centering
\small
\begin{tabular}{lrrrr}
\toprule
\textbf{Metric} & \textbf{SPANN} & \textbf{Starling} & \textbf{DiskANN} & \textbf{GANNS} \\
\midrule
Build time (s)     & 12,113 & 11,432 & 5,095 & \rev{3,199} \\
Disk storage (GB)  & 230.3  & 77.9   & 64.6  & 44.9  \\
\bottomrule
\end{tabular}

\caption{Index construction time and disk storage overhead on DEEP.}
\vspace{-2em}
\label{tab:build-storage-overhead}
\end{table}

\subsection{Index Cost}

\noindent\textbf{Index construction cost.}
\rev{Table~\ref{tab:build-storage-overhead} compares construction time on DEEP~\cite{covington2016deep}. The optimized \projectname{} pipeline overlaps independent local index builds while retaining parallelism within each build, allowing better use of compute resources even with uneven cluster sizes. Each local index contains only $N/n$ vectors on average, where $N=\sum_{i=1}^{n}|C_i|$ and $n$ is the number of disjoint clusters, reducing the scale of each construction task. Together with a lightweight graph abstraction, this enables \projectname{} to complete indexing in \textbf{3{,}199\,s}, \textbf{1.59$\times$} faster than DiskANN~\cite{jayaram2019diskann}.}

\rev{Starling~\cite{wang2024starling} and SPANN~\cite{chen2021spann} incur higher build overhead from their more complex data layouts. SPANN takes \textbf{12{,}113\,s}, as replicated vectors require redundant posting-list construction. By keeping partitions independent and avoiding vector replication, \projectname{} builds \textbf{3.79$\times$} faster than SPANN.}

\noindent\textbf{Disk storage overhead.}
Table~\ref{tab:build-storage-overhead} also compares disk usage on DEEP. \projectname has the smallest footprint at \textbf{44.9\,GB}, including vectors and indices. SPANN requires \textbf{230.3\,GB}, more than \textbf{5$\times$} larger, because vector replication substantially increases storage. DiskANN uses quantized representations and graph metadata to reduce memory pressure, but still needs additional index structures for high recall. Starling is more compact than SPANN, yet its block-aligned layout adds redundancy for locality.

In contrast, \projectname stores each vector once and adds only a compact $\mathcal{GA}$. Even at large scale, the $\mathcal{GA}$ occupies only about \textbf{100\,MB}, negligible compared with the vector store. This duplication-free design keeps storage overhead low while preserving accuracy and out-of-core efficiency.

\subsection{Parameter Sensitivity}\label{sec:eval-parameter}
\begingroup
\emergencystretch=1em

\begin{figure}[t]
  \centering
  \includegraphics[width=\columnwidth]{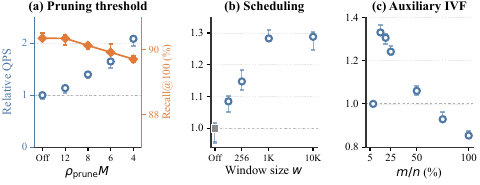}
  \caption{\rev{Parameter sensitivity on TriviaQA. Blue circles show relative QPS; orange diamonds show recall on the right axis in (a). Pruning settings are equally spaced and labeled by $\rho_{\mathrm{prune}}M$; Off disables pruning. QPS is normalized to the disabled configurations in (a,b) and to the lowest plotted auxiliary-IVF coverage in (c).}}
  \label{fig:parameter-sensitivity}
  \Description{\rev{Pruning shows five categorical settings: Off, then products of rho-prune and M equal to 12, 8, 6, and 4, with M=18. Off disables pruning and is the QPS denominator, not a numeric threshold. Blue QPS scatter points use the left axis; orange connected recall diamonds use the right axis. Scheduling uses a logarithmic window axis with a separate Off tick. Auxiliary coverage shows 8, 15, 20, 25, 50, 75, and 100 percent, normalized by same-seed QPS at the unplotted 5 percent reference. Its horizontal ticks still begin at 5 percent. All panels show ranges from three full-query runs.}}
\end{figure}

\rev{Figure~\ref{fig:parameter-sensitivity} evaluates pruning, scheduling, and auxiliary IVF coverage on TriviaQA. In (a), pruning stops after $\lceil\rho_{\mathrm{prune}}M\rceil$ consecutive non-improving clusters ($M=18$). At $\rho_{\mathrm{prune}}M=12$, QPS rises 13.7\% while recall changes from 90.354\% to 90.349\%. Smaller thresholds trade modest recall for fewer visits: $\rho_{\mathrm{prune}}M=4$ yields $2.09\times$ QPS at 89.709\% recall.}

\rev{In (b), scheduling groups queries with similar cluster accesses to improve locality. A window of 1{,}024 raises QPS by 28.3\% without changing median recall; increasing it to 10{,}000 adds only 0.4\% QPS. Larger windows can increase sorting costs and online queueing delays.}

\rev{In (c), $m/n$ denotes the fraction of clusters assigned auxiliary IVF views, selecting the $m$ largest of $n$ clusters. Across the evaluated settings, QPS decreases as auxiliary-IVF coverage expands from 15\% to 100\%, indicating that broader coverage introduces additional list-scanning work that outweighs its benefit for large clusters. Auxiliary IVF should therefore remain targeted at the clusters that dominate memory pressure rather than being applied broadly.}

We also vary the recall target (Recall@1 versus Recall@10) and the number of query threads on DEEP~\cite{covington2016deep}. When moving from Recall@1 to Recall@10, all systems experience a QPS drop of roughly 20\%. Despite this, \projectname{} achieves the highest QPS and lowest latency across the evaluated recall range from 80\% to 98\%.
At Recall@10 $= 90\%$, \projectname delivers \textbf{3.5$\times$} higher QPS than Starling~\cite{wang2024starling}, and at Recall@1 $= 90\%$, it achieves \textbf{3.2$\times$} higher QPS than Starling. In multi-threaded settings, \projectname also scales effectively: with 64 threads, it outperforms Starling by \textbf{2.2$\times$} and SPANN~\cite{chen2021spann} by \textbf{5.4$\times$} in QPS at Recall@10.

\par\endgroup

\subsection{\rev{Incremental Update Evaluation}}
\label{sec:eval-dynamic-update}

\begin{figure}[t]
  \centering
  \includegraphics[width=\columnwidth]{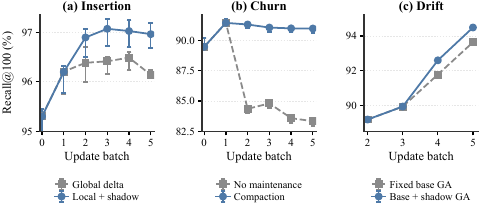}
  \caption{\rev{Recall across update batches (medians and run ranges, three runs). Churn repeatedly inserts and deletes equal-sized update batches. Drift omits the identical batches 0--1 and uses a truncated recall axis.}}
  \label{fig:dynamic-updates}
  \Description{\rev{Three line charts show recall over insertion, churn, and drift batches. Legends below the panels identify update strategies; markers and vertical error bars show measured medians and run ranges.}}
\end{figure}

\rev{Figure~\ref{fig:dynamic-updates} evaluates incremental insertion, balanced churn, and routing drift, using exact Recall@100 at every stage.
For cumulative insertion at the same $efSearch=100$, partition-local deltas with graph-abstraction-guided placement and shadow refresh process the 4.26 million insertions at 19.4K vectors/s, 19.7\% faster than a global delta.
The global-delta baseline instead places all new vectors in one shared delta independent of cluster ownership, yielding higher final query throughput (796.7 versus 439.7 QPS).
However, this shared organization makes compaction difficult to localize and is less suitable for sustained maintenance; it also finishes at 96.14\% recall, compared with 96.97\% for partition-local deltas.}

\rev{The synthetic churn workload first inserts 100K vectors generated by perturbing the query embeddings.
Each of the next four batches inserts 100K new vectors and deletes the preceding batch, creating sustained update turnover while keeping the amount of live data bounded.
Without compaction, final recall falls to 83.33\% and QPS to 174.4.
Localized compaction incorporates accumulated updates into the persistent partition, removes deleted and replaced records, and rebuilds only that partition's local index on SSD.
It maintains 91.00\% recall and 532.6 QPS after 400K deletions, a 3.05$\times$ throughput improvement over no maintenance.
In the drift microbenchmark, allowing the shadow graph abstraction to follow the evolving data distribution raises final recall from 93.64\% to 94.49\% with a 2.6\% QPS cost.}

\rev{Together, the results show that updates can be absorbed through partition-local deltas while maintenance remains confined to the affected partitions.
Localized compaction incorporates accumulated updates into SSD-resident indexes without rebuilding or repartitioning unrelated partitions, while graph-abstraction refresh keeps logical routing aligned with the evolving data distribution.}

\section{Related Work}

While out-of-core ANNS systems are discussed separately in Section~\ref{sec:background}, prior work mainly includes:
(1) algorithmic indexing optimized for low-latency search when the index fits in RAM, and
(2) accelerator-based systems that leverage specialized hardware such as GPUs and PIM to improve throughput.
For billion-scale settings where vectors exceed memory, existing out-of-core designs typically rely on IVF-based indices~\cite{chen2021spann,xu2023spfresh,zhang2024fast,chen2024onesparse}, disk-resident graph indices~\cite{manohar2024parlayann,DBLP:journals/pvldb/FuXWC19}, or hierarchical structures with a small in-memory navigator over SSD-resident data~\cite{chen2021spann,tian2024scalable,jayaram2019diskann}.

\noindent\textbf{Algorithmic ANNS solutions.}
A large body of work studies indexing strategies for in-memory ANNS~\cite{zhu2024gts,gao2024rabitq,voruganti2025mirage,wei2024det,wang2025accelerating,ockerman2025exploring,gao2025practical,gou2025symphonyqg,manohar2024parlayann,zhang2023vbase,rubel2026pipnn}.
HNSW~\cite{malkov2020hnsw} is widely adopted for its multi-layer graph structure, and methods such as SPTAG~\cite{chen2018sptag} combine space partitioning with graph traversal to reduce distance computations.
These systems achieve excellent latency and recall, but they assume that both vectors and index metadata fit in RAM, which limits scalability.

\noindent\textbf{Accelerator-based ANNS solutions.}
Recent systems exploit GPU-based~\cite{groh2022ggnn,yu2022gpu,ootomo2024cagra,zhu2024gts}, PIM-based~\cite{li2025ansmet,wu2025turbocharge,chen2024memanns}, FPGA-based~\cite{Jiang2023codesign}, and CXL-based~\cite{jang2023cxl} designs.
GPUs accelerate graph search and IVF+PQ filtering, but limited device memory constrains dataset size, and CPU--GPU communication can become a bottleneck even with overlapping techniques such as RUMMY~\cite{zhang2024fast}.
Tree-based designs such as GTS~\cite{zhu2024gts} improve GPU utilization but still assume that the working set stays largely in memory.
PIM-based approaches reduce data movement but remain restricted by hardware availability and limited support for complex index structures.

Together, these methods are effective with sufficient DRAM, GPU HBM, or specialized hardware, but do not directly address CPU--SSD search under tight memory budgets, where most vectors and index structures remain SSD-resident.


\section{Conclusion}
We presented \projectname{}, a hierarchical out-of-core ANNS engine that decouples physical locality from logical connectivity.
By keeping vectors in non-replicated SSD partitions, providing boundary reachability through a compact memory-resident graph abstraction, and combining scale-aware local indexes with utility-aware execution, \projectname{} reduces redundant SSD work without sacrificing recall.
\rev{Across five datasets under strict memory constraints, GANNS achieves representative QPS improvements of 2.0--4.7$\times$ while preserving recall.}

\balance
\bibliographystyle{ACM-Reference-Format}
\bibliography{ref}

\clearpage
\appendix
\clearpage

\end{document}